\documentclass[reprint,amsmath,amssymb,aps,physrev]{revtex4-2}
\usepackage{graphicx}
\usepackage{dcolumn}
\usepackage{bm}
\usepackage{algorithm}
\usepackage{algpseudocode}
\usepackage{xcolor}

\begin{document}

\title{Towards Control of High-Frequency Instabilities in Sequential-in-Time \\ Neural PDE Solvers}

\author{Mikhail Kuvakin}
\affiliation{Department of Computer Science,
Tsinghua University, Beijing 100084, China}

\author{Luis Lehner}
\affiliation{Perimeter Institute for Theoretical Physics,
31 Caroline Street North, Waterloo, ON N2L 2Y5, Canada}

\author{Di Luo}
\email{diluo@tsinghua.edu.cn}
\affiliation{Department of Physics,
Tsinghua University, Beijing 100084, China}
\affiliation{Institute of Advanced Study,
Tsinghua University, Beijing 100084, China}

\author{Huan Yang}
\email{hyangdoa@tsinghua.edu.cn}
\affiliation{Department of Astronomy,
Tsinghua University, Beijing 100084, China}

\date{\today}

\begin{abstract}
Uncontrolled growth of high-frequency modes obstructs the numerical evolution of ill-posed differential equations, including those arising in effective field theories and extensions to General Relativity (GR). Conventional regularization modifies the governing equations, restricts the solution space, or filters the numerical representation. In this work, we address this problem  from the perspective of neural PDE solvers by analyzing and extending the Time-Evolving Natural Gradient (TENG) framework. A modal analysis describes how per-step approximation errors seed an ultraviolet (UV) runaway. To mitigate this failure mode, we introduce UV-evading Time-Evolving Natural Gradient (UVNG), which regulates the projection step by penalizing the linearized change in the mean squared spatial derivative. The penalty provides an explicit adjustable control over the tolerance for steep features, while spectral correction fitted to low Fourier modes improves approximation accuracy and accelerates fitting. Across anti--diffusion and Burgers' shock formation, UVNG suppresses high-frequency artifacts and approximately halves the relative error compared with the Physics-Informed Neural Network (PINN) baseline. UVNG also produces stable evolutions of a nonlinear scalar model with pathologies relevant to extensions of GR. Comparison with Israel--Stewart relaxation shows agreement in later pulse propagation and a regulator-dependent difference in pulse-splitting time. These results demonstrate the method's practical ability to regulate instability while offering explicit control over the resolution of steep features within a sequential-in-time neural PDE solver.

\end{abstract}

\maketitle

\section{\label{introduction}Introduction}
Partial differential equations describe many physical phenomena whose nonlinear dynamics must be obtained numerically. Traditional methods, including finite elements and spectral methods, have a long record of success, but ill-posed problems present a different difficulty. By Hadamard's definition~\cite{hadamard}, a problem is ill-posed if a solution does not exist, is not unique, or fails to depend continuously on the initial or boundary data. The last possibility is particularly troublesome for numerical evolution: small errors can grow without a bound that is uniform in resolution. Such pathologies occur in inverse heat conduction and hydrodynamical and gravitational models. In extensions to General Relativity, a common manifestation is an uncontrolled transfer to short wavelengths~\cite{Cayuso2017,Allwright2019}, often referred to as ``UV runaway.'' Here we use this term for the uncontrollable growth of high-frequency oscillations seeded by numerical or approximation errors.

Although ill-posedness limits the physical conclusions that can be drawn from an evolution, these problems arise in physical models and remain worth studying. Regulated evolutions can help identify where a model fails and which features survive different regularization choices. Their dependence on the regulator must therefore be assessed.

Neural PDE solvers provide a complementary way to investigate these difficulties. Physics-Informed Neural Networks (PINNs)~\cite{Raissi2019} fit a solution over spacetime, whereas Time-Evolving Natural Gradient (TENG)~\cite{Luo2024} advances through successive time slices. Global optimization avoids successive propagation of time-slice projection errors, and neural-network spectral bias can discourage fitting high-frequency noise. However, it also has distinct drawbacks, such as lacking temporal causality and having limited scalability due to considerable memory overhead. TENG instead preserves the ordering of the initial-value problem and fits each new slice separately. In an ill-posed regime, its sequential nature makes it vulnerable to the propagation and amplification of high-frequency errors.

In this work, we introduce UV-evading Time-Evolving Natural Gradient (UVNG) to retain the advantages of sequential evolution while providing explicit control within the projection step. Our main contributions are:
\begin{itemize}
    \item We analyze how per-step approximation errors accumulate in TENG, describing how the pathological growth of high-frequency modes unfolds in ill-posed problems.
    \item To control this growth at the projection level, we introduce an adjustable penalty on the linearized change in the mean squared spatial derivative. Furthermore, with correction steps we provide a way to accelerate fitting of low Fourier modes.
    \item On anti--diffusion and Burgers' shock formation, we demonstrate that UVNG suppresses high-frequency artifacts and approximately halves the relative error compared with the PINN baseline.
    \item For a nonlinear model relevant to extensions of GR, UVNG maintains stability and provides later pulse propagation consistent with Israel--Stewart relaxation. Comparing the two methods identifies the regulator dependence of the earlier evolution.
\end{itemize}

\section{\label{background}Background and related work}
Consider an evolution equation, possibly ill-posed, defined on a spatial domain $\mathcal{X} \subset \mathbb{R}^d$ and a temporal domain $\mathcal{T} \subset \mathbb{R}_+$:
\begin{equation}\label{formula:PDE}
\left\{
\begin{aligned}
&\frac{\partial u}{\partial t} = \mathcal{L}\circ u, \quad (t,x) \in \mathcal{T} \times \mathcal{X} \\
&u(0,x) = u_0(x) \\
&\mathcal{B}\left(u(t,x)\right) = 0, \quad x \in \partial \mathcal{X}
\end{aligned}
\right.
\end{equation}
where $\mathcal{L}$ is a spatial differential operator and $\mathcal{B}$ specifies the boundary conditions.

\subsection{Traditional ways to deal with ill-posedness}
Classical approaches to ill-posedness include modification of the governing equation, for example by adding viscosity, quasi-reversibility~\cite{LattesLions1969}, or a damping mechanism~\cite{Israel1979,Cayuso2017,Cayuso:2020lca,Besharat:2026vkg}; restriction of the solution space, for instance through analytic or Gevrey regularity~\cite{Treves1967}; and modification of the numerical representation, such as filtering or stabilization~\cite{GottliebOrszag1977,Boyd2001}. Field redefinitions can also change the UV limit of a theory~\cite{Figueras:2024bba}, although their effect on hyperbolicity requires care~\cite{Thaalba:2026eyv}. These choices can affect the dynamics at resolved scales. Spectral cutoff, for example, restricts the state to a finite set of Fourier modes. With an initially projected state, the filtered evolution can be written as $\partial_t u=\mathcal{P}_k\circ(\mathcal{L}\circ u)$, where $\mathcal{P}_k$ projects onto the retained modes. Such a cutoff suppresses unresolved modes but limits resolution near sharp features and can produce Gibbs oscillations. Smooth filters provide other compromises~\cite{Boyd2001}.

Compared to these methods, UVNG imposes soft regulation of the numerical representation through the solver's projection step. It retains the original equation and supplies an adjustable penalty on derivative changes rather than a fixed spectral cutoff. It therefore shares the need to assess regulator dependence. An ill-posed equation may have a solution for particular smooth data -- as in the anti--diffusion test below -- without admitting a stable initial-value evolution in the chosen function space. Following Refs.~\cite{Cayuso2017,Allwright2019}, in Section~\ref{experiments} we examine whether the regulator suppresses UV growth while preserving the low-frequency dynamics of interest.

\subsection{PINN}
Physics-Informed Neural Networks \citep{Raissi2019} reformulate solving partial differential equations as a global optimization problem. The solution on the spacetime domain is represented by a neural network $u_{\theta}(t,x)$, where the parameters $\theta$ are found by minimizing a composite loss function:
\begin{equation*}
\textit{L}(\theta) = \lambda_{eq} \textit{L}_{eq} + \lambda_{ic} \textit{L}_{ic} + \lambda_{bc} \textit{L}_{bc}
\end{equation*}

Here, $\textit{L}_{eq}$ measures the PDE residual over $\mathcal{T}\times\mathcal{X}$, while $\textit{L}_{ic}$ and $\textit{L}_{bc}$ enforce the initial and boundary conditions. PINNs have been applied across a range of physical problems~\cite{Luo2025,Cao2024,Mao2025,Wang2025,Chomali2026}, with architectural extensions including FBPINNs~\cite{Moseley2023} and separable PINNs~\cite{Cho2023}.

A global spacetime fit does not propagate projection errors through successive time slices. Spectral bias -- the tendency of neural networks to learn low-frequency components first~\cite{Rahaman2019} -- can also limit the fitting of high-frequency noise. These properties distinguish PINNs from sequential-in-time solvers, providing better stability for ill-posed problems in practice. Despite this, PINNs also possess serious drawbacks, which limit their applicability: significant computational costs do not necessarily lead to superior accuracy \citep{Grossmann2024}; solution may suffer from a lack of temporal causality \citep{Mahoney2021, Perdikaris2024}; ill-posedness of the optimization problem becomes severe with higher-order derivatives and constrains the optimization \citep{Rathore2024}. Although the aforementioned issues are addressed \citep{Perdikaris2024, Rathore2024}, they cannot be completely eliminated and limit the performance of PINNs.

\subsection{TENG}
Time-Evolving Natural Gradient \citep{Luo2024} treats the evolution as a sequence of approximations on discrete time slices. Given a solution represented by a neural network $u_{\theta_t}(x)$ at time $t$, TENG finds the parameters $\theta_{t+\Delta t}$ for the subsequent slice by projecting the integrated dynamics into the network's parameter space. A distinguishing feature of the TENG stepper is the two-step update rule. It first takes a gradient step in function space ($u$-space), where the squared-error objective is convex, and then projects that update onto parameter space ($\theta$-space). Formally, given a target function $u_{\text{target}}(x)$ obtained via an explicit temporal integration, the parameters are updated iteratively as follows:
\begin{equation*}
\begin{aligned}
\text{1.}&\text{ Compute function-space update: }\\
&\Delta u(x) = -\alpha \frac{\partial\textit{L}(u_{\theta},u_{\text{target}})}{\partial u_{\theta}} (x)\\
\text{2.}&\text{ Project to parameter space: }\\
&\Delta \theta = \arg \min_{\Delta \theta} \left\| \Delta u(x) - \sum_j \Delta \theta_j\frac{\partial u_{\theta}}{\partial \theta_j}(x) \right\|_{L^2(\mathcal{X})}^2 \\
\text{3.}&\text{ Update parameters: } \theta \leftarrow \theta + \Delta \theta
\end{aligned}
\end{equation*}

Throughout this work, the spatial $L^2(\mathcal{X})$ norm is mean-normalized, with uniform collocation as its quadrature approximation. The second step is solved as a linear least-squares problem. Automatic differentiation evaluates spatial derivatives of the network without a finite-difference stencil. 

Previous work has extended TENG to preserve physical invariants~\cite{ShiWang2026} and to enforce general boundary conditions~\cite{Jiang2026}. Our focus is to control the growth of high-frequency modes. TENG can fit time-step targets accurately in well-posed problems~\cite{Luo2024}. In an ill-posed evolution, however, the target may already contain spurious modes seeded by earlier approximation errors, and an accurate fit can reproduce these modes along with the physical signal. The following analysis describes how approximation errors accumulate in this setting.

\section{\label{theory}Analysis of UV runaway in TENG}
To investigate UV runaway in TENG, we analyze the propagation of approximation errors in a sequential-in-time neural solver. We consider Eq.~\eqref{formula:PDE} on a one-dimensional periodic domain, with a linear constant-coefficient differential operator of order $2n$:
\begin{equation}
\label{formula:L-operator}
\mathcal{L}\circ u = \mathcal{W}\circ u + \lambda(-1)^n \frac{\partial^{2n}u}{\partial x^{2n}}.
\end{equation}
Here $\lambda>0$, and $\mathcal{W}$ contains derivatives of order strictly less than $2n$. Its Fourier symbol is denoted by $W(k)$. For a harmonic mode $e^{ikx}$, the ill-posed term yields $\lambda k^{2n} e^{ikx}$. In the high-frequency regime ($k \gg 1$), this term dominates the dynamics, as its amplification grows polynomially with the wavenumber.

At step $j$, the neural-network approximation $u^j(x)$ introduces a fitting error $\epsilon^j(x)$, which can be decomposed into Fourier modes:
\begin{equation}
\epsilon^j(x) = \sum_m A_j(m)e^{imx}.
\end{equation}
We isolate a mode $k$ and write its contribution as $\epsilon^j(x)=A_j e^{ikx}$. UV runaway occurs because the next time-step target includes and amplifies errors inherited from previous fits. For Euler integration,
\begin{equation}
u^j = \mathcal{E}\circ u^{j-1}+\epsilon^j,
\qquad \mathcal{E}=\mathcal{I}+\Delta t\,\mathcal{L}.
\end{equation}
Each inherited coefficient is multiplied by $1+\Delta t(W(k)+\lambda k^{2n})$. In the UV limit, the leading factor is $G(k)=1+\Delta t\lambda k^{2n}$, with a correction $\Delta t W(k)$. Summing the inherited errors yields the following estimates. The Euler and Heun derivations are given in Appendix~\ref{appendix:proof}.
\begin{widetext}
\newtheorem{proposition}{Proposition}
\begin{proposition}\label{proposition}
Consider the sequential-in-time neural solver applied to the problem \eqref{formula:PDE} with the right hand-side \eqref{formula:L-operator}. Assuming a constant approximation error with amplitude $A$ is injected at each time step, the cumulative error amplitude for mode $k$ at step $N$ is asymptotically given by $\mathcal{A}_{\text{Euler}}(N,k)$ for Euler integration and $\mathcal{A}_{\text{Heun}}(N,k)$ for Heun integration:

\begin{equation}
\label{eq:Euler_formula}
\begin{aligned}
\mathcal{A}_{\text{Euler}}(N,k) =
A\frac{(1+\Delta t \lambda k^{2n})^{N}-1}{\Delta t \lambda k^{2n}} + \mathcal{O}\left(\frac{AN(1+\Delta t \lambda k^{2n})^{N-1}W(k)}{\lambda k^{2n}}\right)
\end{aligned}
\end{equation}
\begin{equation}
\label{eq:Heun_formula}
\begin{aligned}
\mathcal{A}_{\text{Heun}}(N,k) = A_{\text{total}}\frac{\left(1 + \Delta t\lambda k^{2n} + \frac{\Delta t^2 \lambda^2 k^{4n}}{2}\right)^N - 1}{\Delta t\lambda k^{2n} + \frac{\Delta t^2 \lambda^2 k^{4n}}{2}} + &\mathcal{O}\left(\frac{A_{\text{total}} N W(k)}{\lambda k^{2n}}\left(1 + \Delta t\lambda k^{2n} + \frac{\Delta t^2 \lambda^2 k^{4n}}{2}\right)^{N-1}\right),\\
&\text{where}\ A_{\text{total}} = A\left(1 + \frac{\Delta t}{2} (W(k) + \lambda k^{2n})\right).
\end{aligned}
\end{equation}
\end{proposition}
\end{widetext}

Note that the approximation $A_{\mathrm{total}}\approx A$ in Eq.~\eqref{eq:Heun_formula} requires $\Delta t\,|W(k)+\lambda k^{2n}|\ll1$.

\subsection{Extension to multiple dimensions}
In the multidimensional spatial domain $\mathcal{X} \subset \mathbb{R}^d$, the linear differential operator generalizes to:
\begin{equation*}
\mathcal{L}\circ u = \mathcal{W}\circ u + (-1)^n \sum_{|p|=2n} \lambda_{p}\partial_x^{p} u,
\end{equation*}
where $p=(p_1,\dots p_d)$ is a multi-index. For a harmonic mode $\epsilon^j(\boldsymbol{x}) = A_j e^{i(\boldsymbol{k},\boldsymbol{x})}$ corresponding to wave vector $\boldsymbol{k}=(k_1, \dots k_d)$, the leading amplification factors for $\mathcal{E}$ and $\mathcal{H}$, with lower-order corrections handled as in one dimension, are:
\begin{equation}
\label{eq:Euler_new_G}
G(\boldsymbol{k}) = 1 + \Delta t \sum_{|p|=2n} \lambda_{p}(k_1^{p_1}\dots k_d^{p_d}),
\end{equation}
\begin{equation}
\label{eq:Heun_new_G}
\begin{split}
\tilde G(\boldsymbol{k}) = 1 &+ \Delta t\sum_{|p|=2n} \lambda_{p}(k_1^{p_1}\dots k_d^{p_d}) +\\
&+\frac{\Delta t^2 }{2}\left(\sum_{|p|=2n} \lambda_{p}(k_1^{p_1}\dots k_d^{p_d})\right)^2.
\end{split}
\end{equation}

Assuming $\sum_{|p|=2n} \lambda_{p}(k_1^{p_1}\dots k_d^{p_d})>0$, the estimates~\eqref{eq:Euler_formula} and \eqref{eq:Heun_formula} extend accordingly (see Appendix~\ref{appendix:2D_runaway} for more details).
\subsection{Nonlinear operators}
For problems with nonlinear dynamics:
\begin{equation*}
\mathcal{L}\circ u = \mathcal{W}\circ u + \mathcal{N}\circ u + \lambda(-1)^n \frac{\partial^{2n}u}{\partial x^{2n}},
\end{equation*}
where $\mathcal{N}$ is a nonlinear operator. We consider linearization of $\mathcal{N}$ around $u(x)$:
\begin{equation*}
\mathcal{N}\circ (u+\delta u) = \mathcal{N}\circ u + \frac{\partial\mathcal{N}}{\partial u}\circ\delta u + \mathcal{O}\left((\delta u)^2\right),
\end{equation*}
where $\frac{\partial\mathcal{N}}{\partial u}$ is a linear operator with variable coefficients depending on $u(x)$. Under this expansion, the action of the Euler step on a perturbed state takes the form:
\begin{equation*}
\begin{split}
&\mathcal{E}\circ (u+\delta u) = \mathcal{E}\circ u + \delta u + \mathcal{O}\left(\Delta t(\delta u)^2\right) +\\
&+\Delta t \left(\mathcal{W}\circ \delta u + \frac{\partial\mathcal{N}}{\partial u}\circ\delta u + \lambda(-1)^n \frac{\partial^{2n}\delta u}{\partial x^{2n}}\right).
\end{split}
\end{equation*}
If the linearized nonlinear operator has order $m<2n$, with smooth bounded coefficients on the background under consideration, its local high-frequency contribution scales as $\mathcal{O}(k^m)$. The $\lambda k^{2n}$ term therefore dominates the frozen-coefficient symbol. This extends the UV-growth mechanism locally to such nonlinear problems.

\subsection{Validation}
\begin{figure*}
  \centering
  \includegraphics[width=0.8\linewidth]{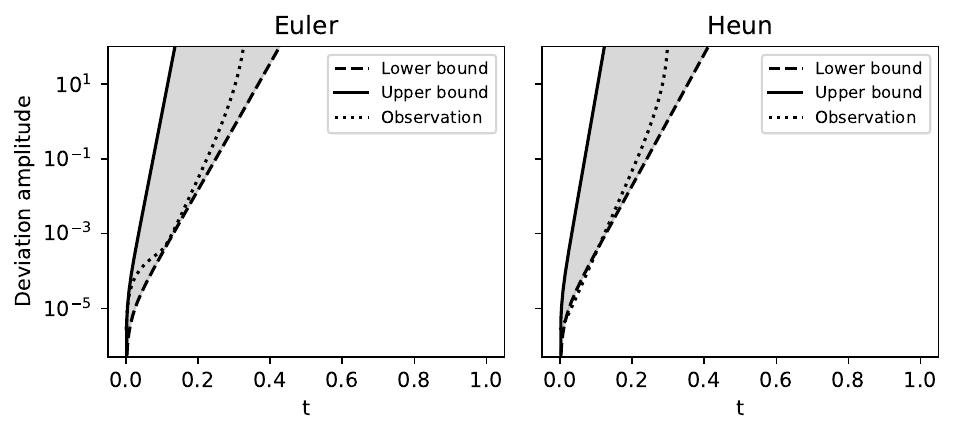}
  \caption{Comparison of accumulated-error estimates and observed TENG deviation for Euler and Heun integration of the anti--diffusion equation~\eqref{eq:AD}.}
  \label{fig:deviation_bounds}
\end{figure*}
The estimates describe how small fitting errors accumulate and amplify at high wavenumber. In the time-resolved limit, with $t=N\Delta t$ fixed and $\Delta t|W(k)+\lambda k^{2n}|\ll1$, the homogeneous amplification approaches $\exp((W(k)+\lambda k^{2n})t)$. The anti--diffusion experiment~\eqref{eq:AD}, for which $W(k)=0$, provides a direct test. Wavenumbers $k\in[9,16]$ dominate the deviation, with per-step error amplitudes in the range $A\in[5\times10^{-7},5\times10^{-6}]$. Under these settings, we can estimate upper and lower bounds for the deviation, and they match the observed behavior (see Fig.~\ref{fig:deviation_bounds}). The structure is familiar from von Neumann analysis; the source here is the error introduced when fitting each neural-network target. Since the ranges of $k$ and $A$ are read off from the run, this checks the consistency of the suggested error accumulation mechanism rather than predicting the error level independently.

This result highlights the existential challenge for TENG in ill-posed regimes: because the stepper seeks to approximate the target function with high fidelity, it treats the exponentially growing high-frequency artifacts as legitimate features of the solution. In the following section we introduce approaches to deal with this issue.

\section{\label{solutions} UV-evading Time-Evolving Natural Gradient}
\subsection{Linear system regularization}
One way to discourage oscillatory neural-network fits is to penalize large parameter values, although this does not directly control their Fourier content. In standard optimization frameworks, this is achieved by introducing a weight decay term directly into the objective function: 
\begin{equation*}
\textit{L}_{\text{reg}} = \textit{L}(u_{\theta}, u_{\text{target}}) + \varepsilon \| \theta \|_2^2.
\end{equation*}
In TENG, the functional gradient is taken with respect to $u_\theta$, so a parameter penalty does not enter that update unless it is treated separately. We instead regularize the projection from the function-space update $\Delta u$ to the parameter increment.
Let $J \in \mathbb{R}^{s \times p}$ denote the Jacobian matrix evaluated at $s$ collocation points for $p$ subsampled parameters, where $J_{ij} = \frac{\partial u_\theta}{\partial \theta_j}(x_i)$. The original TENG projection step solves the linear least-squares problem $J \Delta \theta = \Delta u$. To apply a penalty for significant weight shifts, we introduce $L^2$ regularization directly into the linear system.

While weight regularization postpones the UV runaway, it primarily addresses the symptom of large parameter norms rather than the underlying emergence of pathological high frequencies. To directly tackle the root cause, we introduce a secondary constraint that explicitly penalizes the growth of high-frequency oscillations by limiting the change, at each update, of the mean squared $n$-th derivative of the solution (with $n=2$ in all experiments) \footnote{One may also think of normalizing the value: $\Omega_{\text{norm}}(u_\theta) = \frac{\Omega(u_\theta)}{\mathbb{E}_{x\in\mathcal{X}}\left[ u^2_\theta\right]}$ to avoid collateral penalization of high-amplitude signals.}:
\begin{equation}
\Omega(u_\theta) = \mathbb{E}_{x\in\mathcal{X}}\left[\left(\frac{\partial^n u_\theta}{\partial x^n}\right)^2\right].
\end{equation}
The resulting UV-evading projection step is formulated as an augmented linear system:
\begin{equation}\label{formula:UVNG_system}
\begin{pmatrix} J \\ \gamma I_p \\ \Upsilon \nabla_\theta \Omega(u_\theta)^T \end{pmatrix} \Delta \theta = \begin{pmatrix} \Delta u \\ \textbf{0}_p \\ 0 \end{pmatrix}.
\end{equation}

The augmented system is solved in the least-squares sense. Here $I_p\in\mathbb{R}^{p\times p}$ is the identity matrix, $\gamma$ regularizes parameter increments, and $\Upsilon$ penalizes the linearized change of $\Omega$. Correspondingly, the updated optimization problem has the form:
\begin{widetext}
\begin{equation}
\Delta \theta = \arg \min_{\Delta \theta} \left(s\left\| \Delta u(x) - \sum_j \Delta \theta_j\frac{\partial u_{\theta}}{\partial \theta_j}(x) \right\|_{L^2(\mathcal{X})}^2 + \gamma^2\left\|\Delta \theta\right\|^2_2 + \Upsilon^2\left(\sum_j \Delta \theta_j\frac{\partial \Omega(u_\theta)}{\partial \theta_j}\right)^2\right).
\end{equation}
\end{widetext}

The final term penalizes $(\nabla_\theta\Omega\cdot\Delta\theta)^2$, the square of the first-order change in $\Omega$. Thus $\Upsilon$ sets an explicit trade-off between fitting the time-step target and changing its derivative content. Unlike a penalty on the absolute magnitude of derivatives, UVNG acts on their change during each update. It can therefore retain existing sharp structures while limiting further steepening.

Under certain simplifying assumptions, the regularized projection can also be interpreted as a modification of the governing equation by a term proportional to $\partial^{2n}u/\partial x^{2n}$. This term should not be read as a viscosity: its coefficient is a global, solution-dependent functional that can take either sign. This interpretation is formalized in the following proposition, with the derivation provided in Appendix~\ref{appendix:interpretation}.

\begin{widetext}
\begin{proposition}\label{proposition-2} 
Consider the sequential-in-time neural solver applied to the problem \eqref{formula:PDE} with the regularization imposed on the mean squared $n$-th derivative of the solution ($\Omega(u_{\theta})$). Neglecting parameter-increment regularization ($\gamma=0$), and assuming a tangent space that spans the relevant function-space updates, a smooth periodic solution, and a single regularized projection with $\Delta u=\Delta t\,\mathcal{L}\circ u$ per time step, the resulting dynamics are formally described, in the limit $\Delta t\rightarrow 0$, by the modified equation:
\begin{equation}
\frac{\partial u}{\partial t} = \mathcal{L}\circ u - \nu(u) \frac{\partial^{2n} u}{\partial x^{2n}},\ \text{where}\ \nu(u)=\frac{4\Upsilon_{\text{eff}}^2 \langle\frac{\partial^{2n}u}{\partial x^{2n}},\mathcal{L}\circ u\rangle}{1 + 4\Upsilon_{\text{eff}}^2 \left\|\frac{\partial^{2n}u}{\partial x^{2n}}\right\|^2},\ \Upsilon_{\text{eff}}=\frac{\Upsilon}{\sqrt{s}}.
\end{equation}
\end{proposition}
\end{widetext}

The structure of this result is clearest in the limit $\Upsilon\to\infty$, where, for $\partial_x^{2n}u\neq0$, $\nu(u)\to\langle\partial^{2n}_x u,\mathcal{L}\circ u\rangle/\|\partial^{2n}_x u\|^2$ and the modified equation becomes the projection of the original dynamics onto the surface $d\Omega/dt=0$. The regularized projection thus softly enforces that each step leaves $\Omega$ unchanged; note that it penalizes decreases of $\Omega$ as well as increases.

Proposition~\ref{proposition-2} describes a single regularized projection per time step and therefore does not fully characterize the stabilizing mechanism of UVNG, for two reasons. First, in practice the projection is conducted for a small set of subsampled parameters and iterated many times per time step (Algorithm~\ref{algorithm:UVNG}). Each iteration again fits the remaining residual, so the penalty does not remove the component of the target along $\delta\Omega/\delta u$; it only slows its convergence, the number of iterations needed along that direction growing like $1+\Upsilon_{\text{eff}}^2\|\delta\Omega/\delta u\|^2$, which is large precisely for high-frequency content. Second, the spectral correction step introduced below refits the low Fourier modes of the residual without any penalty. The effective strength of the regulator is therefore set jointly by $\Upsilon_{\text{eff}}$, the number of iterations per time step and the spectral correction, if applied. The distinction matters in practice. For a single mode $u=a\sin kx$ of the anti--diffusion equation~\eqref{eq:AD} with $n=2$, $\partial^4_x u$ is parallel to both $u$ and $\mathcal{L}\circ u$, and the proposition gives $\partial_t u = \lambda k^2 u/(1+4\Upsilon_{\text{eff}}^2k^8\|u\|^2)$; that is, the penalty works against the physical growth. The resulting bias is visible without correction and is reduced when spectral correction refits the physical low frequency mode (Fig.~\ref{fig:AD_ablation}, right). Consistently, directly adding the corresponding higher-order term to the equations does not completely reproduce the behavior of UVNG. We further investigate this in Sec.~\ref{exp:EFT}, where we compare UVNG with a finite-difference solution of the system augmented by the corresponding higher-order correction.

\subsection{Correction steps}
\begin{figure*}
  \centering
  \includegraphics[width=0.8\linewidth]{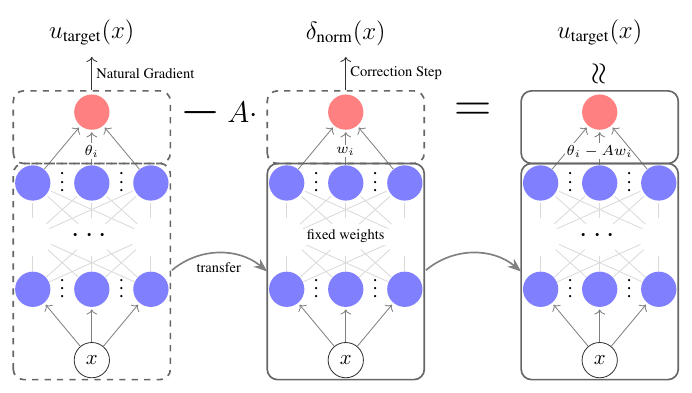}
  \caption{Schematic overview of the correction step. The final model is interpreted as a superposition of two functional components sharing identical latent features. The primary component (left) captures the main dynamics of $u_{\text{target}}(x)$ by the regularized Natural Gradient projection. The auxiliary component (center) functions as a linear corrector, approximating normalized residual $\delta_{\text{norm}}(x)$.}
  \label{fig:uvng_scheme}
\end{figure*}
\subsubsection{Linear correction}
While the regularized projection aims to mitigate UV runaways, the additional constraints inherently restrict the expressivity of the update step and increase computational complexity. To recover approximation speed without incurring memory overhead, we propose a post-projection correction step. We draw inspiration from the Random Feature Method \citep{Chen2022}, which treats a neural network as a linear combination of the activations from its final hidden layer:
\begin{equation*}
u_\theta(x)=b+\sum_{i=1}^l w_i\Phi_i(x),
\end{equation*}
where $\Phi_i(x)$ are the final hidden-layer activations and $\{w_i,b\}$ are the output weights and bias. Correcting the residual is then a linear least-squares problem. Let $\delta(x)=u_\theta(x)-u_{\mathrm{target}}(x)$, $A=\max_x|\delta(x)|$, and $\delta_{\mathrm{norm}}(x)=\delta(x)/A$. If $A=0$, no correction is needed. Let $\Phi\in\mathbb{R}^{s\times(l+1)}$ contain the final-layer activations and the bias column at the collocation points. A naive linear correction step would directly fit the final layer to the residual by solving the system:
\begin{equation}
\Phi\cdot\boldsymbol{w}=\boldsymbol{\delta}_{\mathrm{norm}}
\end{equation}
Applying the resulting update $\boldsymbol{w}$ to existing weights $\boldsymbol{\theta}_{\text{last layer}}$ would eliminate the residual, yielding $u_{\theta_{\text{new}}}(x) \rightarrow u_{\text{target}}(x)$. Fig.~\ref{fig:uvng_scheme} offers a schematic intuition of the approach.\\
However, this unconstrained fitting reintroduces the exact high-frequency noise we previously sought to avoid, negating the benefits of the regularized projection step.

\subsubsection{Spectral correction}
To accelerate fitting without requiring the correction to reproduce high-frequency errors, we aim to fit only selected low Fourier modes of the residual. Choose modes $\{m_1,\ldots,m_d\}$ and apply the same truncated Fourier transform to the feature matrix and the normalized residual. Let $T\in\mathbb{R}^{2d\times s}$ have rows corresponding to the sine and cosine bases of those modes evaluated at the collocation points. The spectral correction step is then formulated as:
\begin{equation}
\label{formula:SC_system}
T\cdot\Phi\cdot\boldsymbol{w}=T\cdot\boldsymbol{\delta}_{\mathrm{norm}}
\end{equation}
By solving this transformed system, we obtain a correction $\boldsymbol{w}$ that optimally adjusts the final layer's weights to eliminate large-scale, low-frequency errors while remaining strictly blind to high-frequency artifacts; these pathological modes are instead continuously regulated by the initial projection step (Eq.~\eqref{formula:UVNG_system}). Algorithm~\ref{algorithm:UVNG} summarizes the combined procedure, including acceptance of a correction only when it lowers the full spatial fitting error.

\begin{figure*}[tbp]
\begin{minipage}{\linewidth}
\begin{algorithm}[H]
\caption{\text{UVNG stepper}}
\label{algorithm:UVNG}
\begin{algorithmic}
[1]\Require Initial parameters $\theta_{\text{init}}$, target function $u_{\text{target}}$, number of global iterations $N_g$, local iterations $N_\ell$, subsample size $p$, regularization coefficients $\gamma, \Upsilon$, number of spectral modes $d$\State $\theta \gets \theta_{\text{init}}$
\For{$k = 1, \dots, N_g$}
\For{$\ell = 1, \dots, N_\ell$}
\State Sample parameter subset $\mathcal{I} \subset \{1,\dots,|\theta|\}$ where $|\mathcal{I}| = p$
\State Evaluate Jacobian $J$ and curvature penalty $\nabla_\theta \Omega(u_\theta)$ on $\mathcal{I}$
\State Compute function-space update $\Delta u$
\State Solve the regularized system (Eq. \ref{formula:UVNG_system}) for $\Delta \theta_{\mathcal{I}}$
\State Update parameters: $\theta_{\mathcal{I}} \gets \theta_{\mathcal{I}} + \Delta \theta_{\mathcal{I}}$
\EndFor
\State Compute residual $\delta(x)=u_\theta(x)-u_{\mathrm{target}}(x)$ and $A=\max|\delta(x)|$
\State Normalize: $\delta_{\mathrm{norm}}(x)=\delta(x)/A$
\State Evaluate final-layer feature matrix $\Phi$ and Fourier projection matrix $T$
\State Solve the spectral correction system (Eq. \ref{formula:SC_system}) for $\boldsymbol{w}$
\State Form candidate $\theta_{\mathrm{new}}$ with $\theta_{\mathrm{last\ layer}}^{\mathrm{new}}=\theta_{\mathrm{last\ layer}}-A\boldsymbol{w}$
\If{$\left\|u_{\theta_{\text{new}}}(x)-u_{\text{target}}(x)\right\|_{L^2(\mathcal{X})}^2 < \left\|u_{\theta}(x)-u_{\text{target}}(x)\right\|_{L^2(\mathcal{X})}^2$}
        \State Update final-layer parameters: $\theta_{\text{last layer}} \gets \theta_{\text{last layer}} - A\boldsymbol{w}$
    \EndIf
\EndFor
\State 
\Return $\theta$
\end{algorithmic}
\end{algorithm}
\end{minipage}
\end{figure*}

To validate our design choices, Appendix~\ref{appendix:ablation} presents an ablation study analyzing the proposed correction steps (Appendix~\ref{appendix:step_ablation}) and illustrating how each component of UVNG contributes to evading UV runaway (Appendix~\ref{appendix:AD_ablation}).

\section{\label{experiments} Experiments}
Sections~\ref{exp:AD} and \ref{exp:Burgers} examine two different difficulties: exponential UV growth in anti--diffusion and loss of smoothness in Burgers' shock formation. We compare UVNG with the baselines specified below. To quantify accuracy, we evaluate the relative $L^2$ error against reference solutions (descriptions of the solutions are provided in Appendix \ref{appendix:reference_solutions}):
\begin{equation*}
\mathrm{E}_{\text{rel}} = \frac{\|u_{\text{pred}}-u_{\text{truth}}\|_{L^2(\mathcal{T} \times \mathcal{X})}}{\|u_{\text{truth}}\|_{L^2(\mathcal{T} \times \mathcal{X})}}
\end{equation*}
To represent global-in-time approaches, we employ the original PINN framework, augmented with causal training \citep{Perdikaris2024} and residual-based adaptive point generation. For a fair comparison, we refrain from introducing specialized ansatz modifications and instead use multi-layer perceptrons (MLPs) for both UVNG and PINN. Regarding sequential-in-time methods, we consider Finite Difference schemes, the original TENG framework and the Optimization-Based Time Integration method (OBTI) \citep{Luo2024}. A comprehensive description of all methods, accompanied by a detailed account of the employed hyperparameters, is provided in Appendix \ref{appendix:exp_setup}.

Additionally, section~\ref{exp:EFT} applies UVNG to a nonlinear scalar model with pathologies relevant to extensions of GR. There, Israel--Stewart relaxation provides a second regulator for comparison rather than a reference solution of known accuracy.

All experiments were conducted on an NVIDIA GeForce RTX 2080 Ti (11 GB VRAM) and an Intel Xeon Silver 4210 CPU.

\subsection{\label{exp:AD} Anti--diffusion equation}
\begin{figure*}
  \centering
  \includegraphics[width=0.85\linewidth]{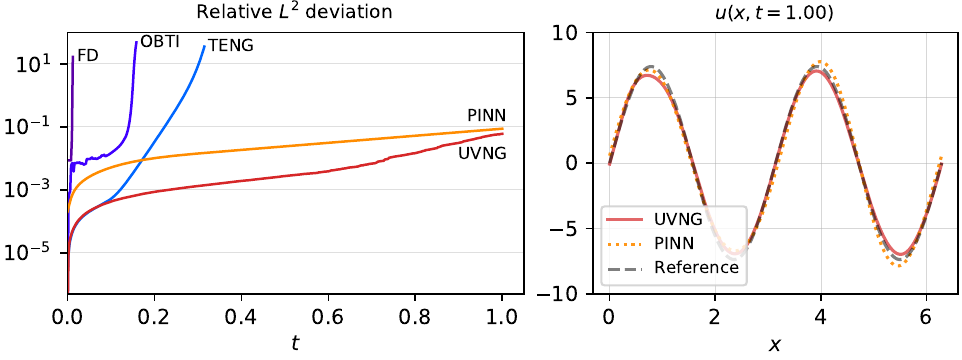}
  \caption{Performance comparison on anti--diffusion equation \eqref{eq:AD}. \textbf{(Left)} Temporal evolution of the relative $L^2$ error demonstrates that only PINN and UVNG prevent high-frequency amplification over the simulated interval. \textbf{(Right)} Spatial profiles of the solutions at the final time step ($t=1$).}
  \label{fig:AD_methods_comparison}
\end{figure*}
The anti--diffusion equation is a standard example of Hadamard instability, arising in inverse heat conduction. Its negative diffusivity amplifies high-frequency errors at a rate that grows without bound with wavenumber. We consider
\begin{equation}
\label{eq:AD}
\left\{\begin{aligned}
&\frac{\partial u}{\partial t}=-\lambda u_{xx}\\
&u(0,x)=A\sin(kx)\\
&u(t,0)=u(t,2\pi),\quad u_x(t,0)=u_x(t,2\pi)
\end{aligned}\right.
\end{equation}
with $\lambda=1/2$, $A=1$, and $k=2$. In this severely ill-posed setting, UVNG and PINN are the only methods that avoid UV runaway while preserving meaningful features of the solution, with UVNG achieving higher accuracy in this setting (see Fig.~\ref{fig:AD_methods_comparison} and Table~\ref{tab:results}). We note that for this single-mode initial datum a Fourier method with a low-pass filter retaining the mode $k=2$ would be essentially exact. The purpose of this test is to expose the failure mode of sequential-in-time neural solvers analyzed in Sec.~\ref{theory} and to show that UVNG removes it, not to compete with spectral methods on a problem tailored to them.

\subsection{\label{exp:Burgers} Burgers equation (inviscid limit)}
\begin{figure*}
  \centering
  \includegraphics[width=1.0\linewidth]{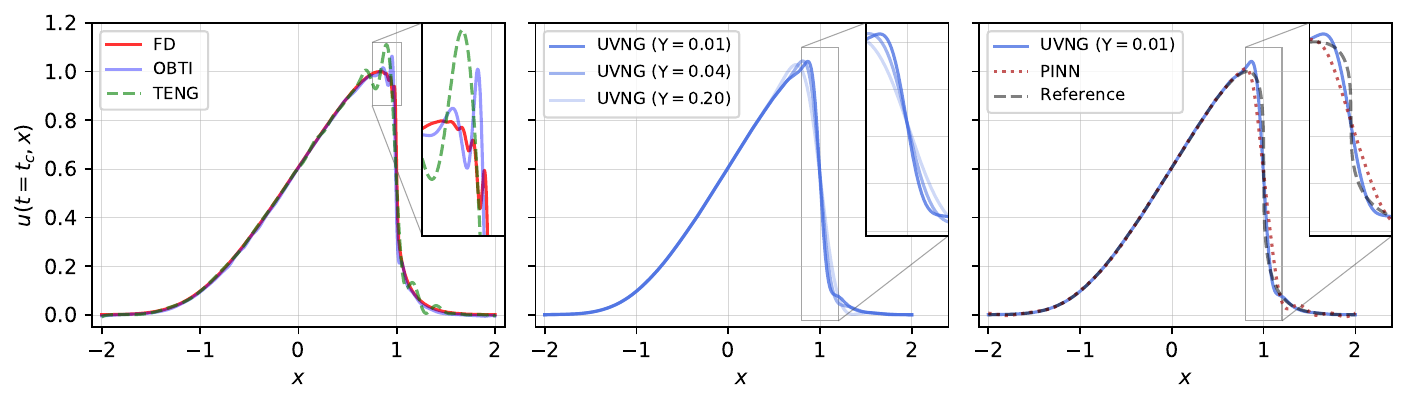}
  \caption{Shock formation analysis in the Burgers' equation \eqref{eq:Burgers} at $t=t_c$. \textbf{(Left)} Spurious oscillations in solutions by alternative methods. \textbf{(Center)} Impact of the hyperparameter $\Upsilon$ on shock resolution illustrates how UVNG can explicitly regulate tolerance to emerging high-gradient features. \textbf{(Right)} Comparison of the predicted spatial profiles with the reference solution.}
  \label{fig:Burgers_shock_comparison}
\end{figure*}

\begin{figure*}
  \centering
  \includegraphics[width=0.9\linewidth]{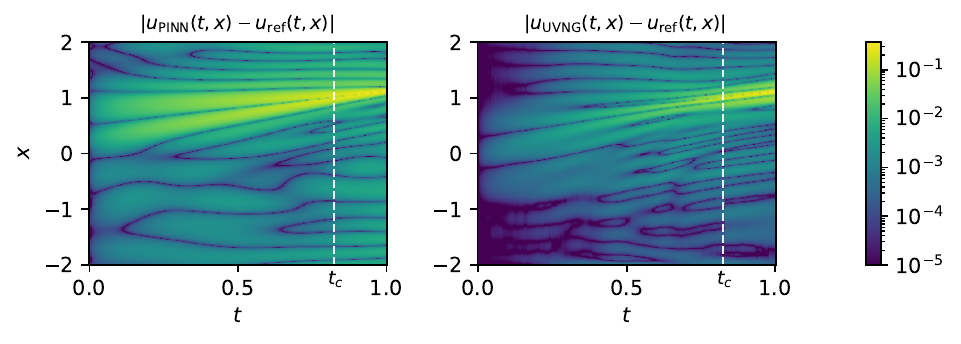}
  \caption{Spatiotemporal error maps illustrating the absolute deviation from the reference solution for PINN \textbf{(Left)} and UVNG \textbf{(Right)}.}
  \label{fig:Burgers_error_colormap}
\end{figure*}
We next consider shock formation in Burgers' equation. Smooth solutions develop a discontinuity, after which a weak formulation and an entropy condition select the physical solution~\cite{leveque2002finite,toro2009riemann}. This differs from the ill-posed anti--diffusion problem: Burgers' equation remains well-posed in the entropy-solution framework, although a smooth neural approximation must resolve a narrowing transition.
\begin{equation}
\label{eq:Burgers}
\left\{\begin{aligned}
&\frac{\partial u}{\partial t} = -\frac{\partial}{\partial x}\left(\frac{u^2}{2}\right)\\
&u(0,x) = e^{-2x^2}\\
&u(t,-2) = u(t,2)
\end{aligned}\right.
\end{equation}
The initial data form a shock at $(t_c,x_c)=(\sqrt{e}/2,1)$. The centered FD, OBTI, and TENG implementations reproduce the early evolution but develop oscillations near the steepening front (Fig.~\ref{fig:Burgers_shock_comparison}). Meanwhile, PINN and UVNG suppress these oscillations. UVNG additionally provides explicit control over the resolved transition width through $\Upsilon$, allowing the resolution of steep features to be adjusted within the projection step, resulting in higher accuracy (see Fig.~\ref{fig:Burgers_error_colormap} and Table~\ref{tab:results}). Appendix~\ref{appendix:Burgers-shock} shows that the UVNG shock trajectory closely follows both the viscous reference and the Rankine--Hugoniot trajectory reconstructed from it. This supports the accuracy of the front motion despite the finite transition width. Furthermore, our results reveal the robustness of UVNG, which remains invariant, up to numerical precision, to whether the advection term is defined in conservative or non-conservative form (see Appendix~\ref{appendix:Burgers-advection}).

\begin{table*}
  \caption{\label{tab:results} Comparison of $\mathrm{E}_{\text{rel}}$ for solutions by considered methods. Each numerical entry is reported as mean ($\pm$) standard deviation over 3 runs. In case of a failure, the reason is noted as follows: \textit{Runaway} denotes high-frequency growth in anti--diffusion; \textit{Oscillatory} denotes spurious oscillations near the Burgers front.}
  \centering
  \begin{ruledtabular}
  \begin{tabular}{lccccc}
    {\textbf{Problem} \textbackslash \textbf{Method}} & \textbf{FD} & \textbf{OBTI} & \textbf{TENG} & \textbf{PINN} & \textbf{UVNG} \\ \hline
    Anti--diffusion & \textit{Runaway} & \textit{Runaway} & \textit{Runaway} & $6.2\times10^{-2}$ $\pm$ $8.0\times10^{-3}$ & $2.9\times10^{-2}$ $\pm$ $2.5\times10^{-3}$ \\
    Shock formation & \textit{Oscillatory} & \textit{Oscillatory} & \textit{Oscillatory} & $5.7\times10^{-2}$ $\pm$ $2.8\times10^{-3}$ & $3.0\times10^{-2}$ $\pm$ $1.2\times10^{-4}$\\
  \end{tabular}
  \end{ruledtabular}
\end{table*}

\subsection{\label{exp:EFT} Toy model for extensions to GR}
\begin{figure*}
  \centering
  \includegraphics[width=0.75\linewidth]{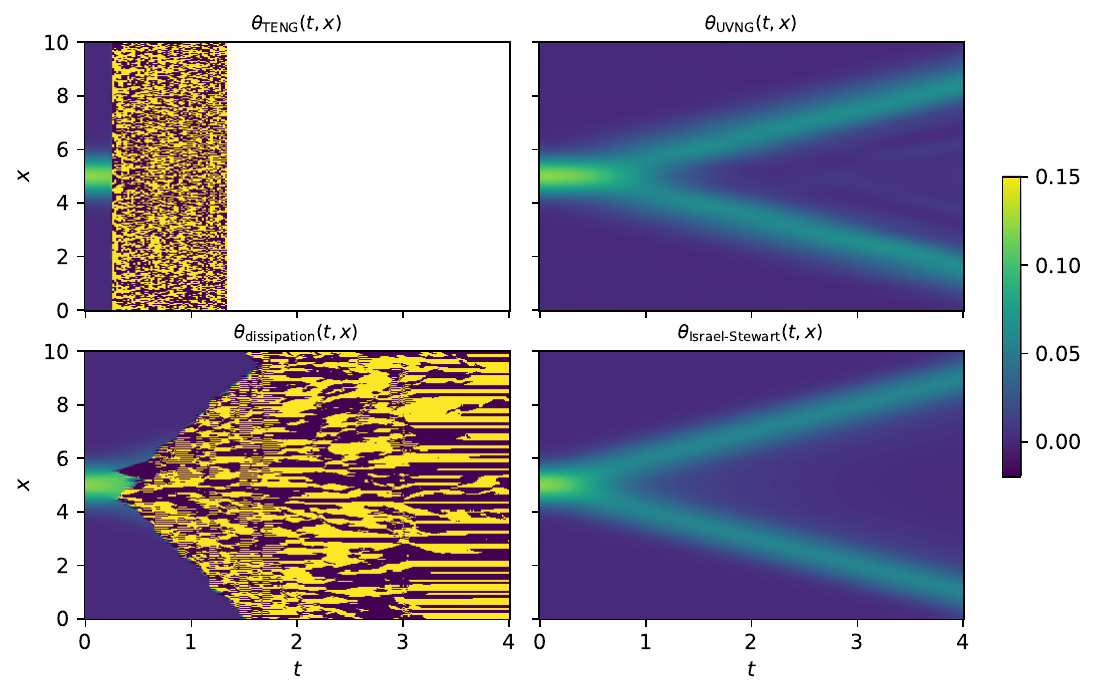}
  \caption{Spacetime evolution of the scalar field $\theta(t,x)$ for the model problem \eqref{eq:EFT} provided by TENG \textbf{(Top left)} and UVNG \textbf{(Top right)}, compared with direct integration of the system augmented with the higher-order correction motivated by Proposition~\ref{proposition-2} \textbf{(Bottom left)} and the system \eqref{eq:EFT_fixed} stabilized by the Israel--Stewart approach \textbf{(Bottom right)}. The TENG evolution terminates after UV runaway; the remaining blank portion of the top-left panel contains no solution data.}
  \label{fig:EFT_theta_colormap}
\end{figure*}

\begin{figure*}
  \centering
  \includegraphics[width=1.0\linewidth]{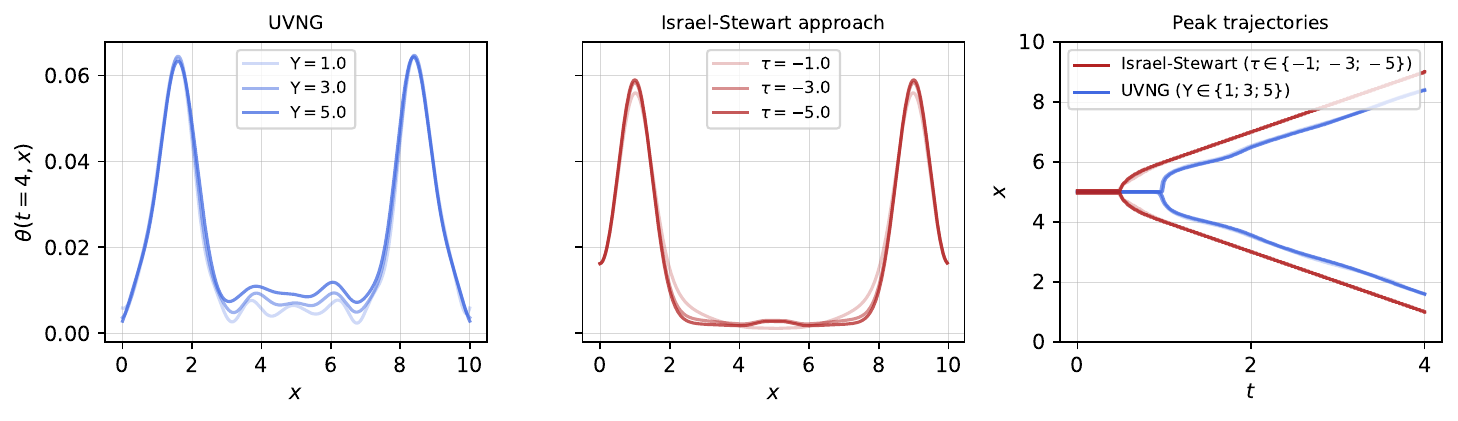}
  \caption{Solutions at $t=4.0$ provided by UVNG (varying $\Upsilon$) \textbf{(Left)} and the Israel--Stewart approach (varying $\tau$) \textbf{(Center)}. Peak trajectories tracking the spatial position of the pulses over time \textbf{(Right)}.}
  \label{fig:EFT_theta_comparison}
\end{figure*}

Extensions of general relativity, particularly those formulated as truncated effective field theories, can introduce higher-derivative interactions (e.g. \cite{Yang:2026erj,Yang:2026csy}) that spoil strong hyperbolicity and/or induce crossing of characteristics and may render the associated initial-value problem ill-posed. To investigate these difficulties in a controlled setting, we consider a scalar-field model originally introduced by Burgess and Williams \cite{Burgess:2014lwa}, which has subsequently been used as a test bench for regularization strategies \cite{Allwright2019,Figueras:2025wtx}. We evaluate our approach on the one-dimensional $\mathcal{O}(M^{-2})$ truncation of this model, following \cite{Allwright2019}. Defining the variables $f\equiv\theta_x$ and $g\equiv\theta_t$, we write the system as the following coupled first-order equations:

\begin{equation}
\label{eq:EFT}
\left\{\begin{aligned}
&\frac{\partial f}{\partial t} = g_x\\
&\frac{\partial g}{\partial t} = f_x \left(1 - \frac{4(f^2 + g^2)}{M^2 - 2f^2 + 6g^2}\right) + g_x \left(\frac{8 f g}{M^2 - 2f^2 + 6g^2} \right)\\
&\frac{\partial \theta}{\partial t} = g
\end{aligned}\right.
\end{equation}
We impose periodic boundary conditions over the spatial domain $x\in[0,10]$, with initial data
\begin{equation*}
\left\{\begin{aligned}
&f(0,x) = -4A(x-5)e^{-2(x-5)^2}\\
&g(0,x) = 0\\
&\theta(0,x) = Ae^{-2(x-5)^2}
\end{aligned}\right.
\end{equation*}
where $M=1/4$ and $A=0.12$. 

Field-dependent coefficients enter the principal part, allowing steepening and, in some regimes, a change of PDE type. To locate the initial data relative to the hyperbolic regime, we write the principal part of Eq.~\eqref{eq:EFT} as $\partial_t g = a(f,g)f_x + b(f,g)g_x$. The eigenvalues $\mu$ of the spatial principal matrix satisfy $\mu^2 - b\mu - a = 0$, so the system is hyperbolic where $b^2+4a>0$. On the initial-data branch $g=0$ and $|f|<M/\sqrt{2}$, this condition reduces to $a=1-4f^2/(M^2-2f^2)>0$, or $|f|<M/\sqrt{6}\approx0.102$. The initial data reach $\max|f| = 2Ae^{-1/2}\approx 0.146$, where $a\approx -3.2$; the system therefore starts out elliptic in two bands, $0.24 \lesssim |x-5| \lesssim 0.82$, around the center of the pulse. There the $\mathcal{O}(M^{-2})$ correction is not small compared to the leading term, so the truncated theory is evaluated outside its regime of validity and the early evolution is determined by whichever regulator is used.

Among the regularization strategies assessed for this model in Ref.~\cite{Allwright2019}, we use the approach introduced in Ref.~\cite{Cayuso2017} for comparison. It is inspired by Israel--Stewart relativistic hydrodynamics~\cite{Israel1979}, and we refer to it as Israel--Stewart relaxation. Consider a generic evolution equation for a metric tensor $g$:
\begin{equation*}
G\circ g = \lambda S\circ g.
\end{equation*}
The approach alters this system by introducing an auxiliary dynamical variable $\Pi$ and a parameter $\tau$:
\begin{equation*}
\left\{\begin{aligned}
&G\circ g = \Pi\\
&\frac{\partial\Pi}{\partial t} = \frac{1}{\tau} (\Pi-\lambda S\circ g)
\end{aligned}\right.
\end{equation*}
With this sign convention, $\tau<0$ drives $\Pi$ toward the source $\lambda S\circ g$ on a timescale $|\tau|$. Applied to the system \eqref{eq:EFT}, the Israel--Stewart approach yields the modified formulation:
\begin{equation}
\label{eq:EFT_fixed}
\left\{\begin{aligned}
&\frac{\partial f}{\partial t} = g_x\\
&\frac{\partial g}{\partial t} = f_x - \Pi\\
&\frac{\partial \theta}{\partial t} = g\\
&\frac{\partial \Pi}{\partial t} = \frac{1}{\tau}\left( \Pi - \frac{2}{M^2}\left( (f^2 - g^2)\Pi +\right.\right.\\
&\qquad\qquad\left.\left.+ 2(f_{x}f^2 + (f_x - \Pi)g^2 - 2g_{x}fg)\right) \right)
\end{aligned}\right.
\end{equation}
Setting the right-hand side of the $\Pi$ equation to zero recovers the algebraic source in Eq.~\eqref{eq:EFT}.

\begin{figure}[tbp]
  \centering
  \includegraphics[width=1.0\linewidth]{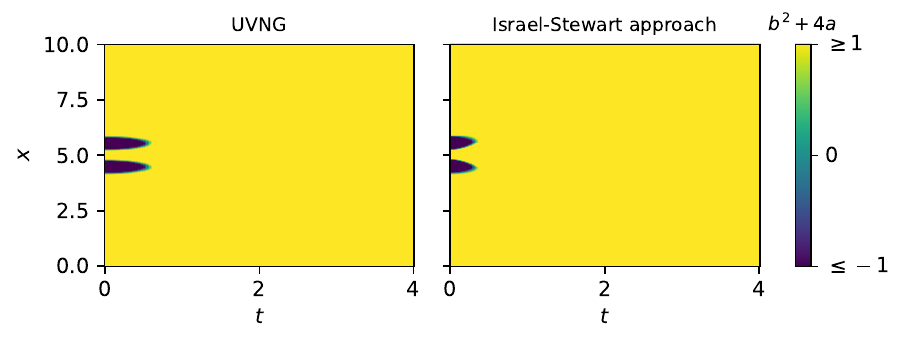}
  \caption{Spatiotemporal map of the characteristic discriminant $b^2+4a$ of the original system \eqref{eq:EFT}, evaluated on the solutions provided by UVNG \textbf{(Left)} and Israel--Stewart approach \textbf{(Right)}.}
  \label{fig:EFT_hyperbolicity}
\end{figure}

In this experiment, we compare TENG and UVNG evolutions of the original system~\eqref{eq:EFT} with the Israel--Stewart stabilized system~\eqref{eq:EFT_fixed}. As shown in Fig.~\ref{fig:EFT_theta_colormap}, the baseline TENG method rapidly develops high-frequency numerical noise and encounters UV runaway. Meanwhile UVNG maintains a stable, smooth evolution over the simulated interval. UVNG and Israel--Stewart relaxation provide similar pulse profiles and consistent later propagation speeds, although the initial pulse splits later in UVNG (Fig.~\ref{fig:EFT_theta_comparison}, right). The hyperbolicity analysis helps explain why the splitting time differs between methods even though their later propagation speeds agree. Fig.~\ref{fig:EFT_hyperbolicity} shows the discriminant of the original system evaluated on each regulated evolution. The early evolution includes regions where the original system is elliptic, and the two regulators give different splitting times. As the pulses separate and decay, the original system evaluated on the regulated fields returns to the hyperbolic regime, where the later propagation speeds agree.

The two regulators act on different parts of the dynamics. In the Israel--Stewart approach, $\tau$ sets the relaxation timescale of the auxiliary field. In contrast, UVNG retains formulation~\eqref{eq:EFT} and uses $\Upsilon$ to control changes in derivative content during the projection of all three fields. These differences are visible in the pulse profiles in Fig.~\ref{fig:EFT_theta_comparison} (left and center). Varying $\tau$ affects the relaxed wave amplitudes, while varying $\Upsilon$ controls the secondary oscillations in UVNG at the fitting budget used here. Within each method, the peak trajectories overlap almost completely across the tested parameter values (Fig.~\ref{fig:EFT_theta_comparison}, right).

To examine whether the stabilization achieved by UVNG can be attributed to the effective higher-order correction identified in Proposition~\ref{proposition-2}, we additionally solve the original system augmented with this correction in the evolution equation for $g$, using the solution-dependent coefficient $\nu(g)$. The correction is applied exclusively to this equation, while the remaining equations, $f_t=g_x$ and $\theta_t=g$, are left unchanged to preserve the compatibility relations associated with the definitions $f=\theta_x$ and $g=\theta_t$. Although this modification delays the development of UV runaway, it does not prevent eventual instability (Fig.~\ref{fig:EFT_theta_colormap}, bottom left). This failure should not be over-interpreted: whenever $\nu<0$ the added term is itself an anti-diffusive fourth-order term, so the augmented system is ill-posed at those times. The test mainly confirms that Proposition~\ref{proposition-2}, which describes a single regularized projection, does not capture the iterated sub-sampled projection composed with weight regularization.

In the UVNG evolution, the independent fits of $f$, $g$, and $\theta$ motivate a direct check of the compatibility conditions $f=\theta_x$ and $g=\theta_t$ during the evolution. The corresponding relative violations
\begin{equation*}
\frac{\|f-\partial_x\theta\|_{L^2(\mathcal{X})}}{\|f\|_{L^2(\mathcal{X})}},\quad
\frac{\|g-\partial_t\theta\|_{L^2(\mathcal{X})}}{\|g\|_{L^2(\mathcal{X})}}
\end{equation*}
remained below $0.3$ throughout the whole simulation $t\in(0,4]$.

\section{\label{limitations} Conclusion and future work}

In this work, we addressed the longstanding challenge of uncontrolled amplification of high-frequency modes in ill-posed problems from the perspective of neural PDE solvers. The modal analysis explains how fitting errors seed high-frequency growth in TENG. To mitigate this failure mode, we have introduced UV-evading Time-Evolving Natural Gradient, which controls the linearized change in the mean squared spatial derivative of the solution during projection through an adjustable penalty, while correction steps fitted to low Fourier modes improve approximation accuracy and fitting speed. This combination preserves temporal causality and supplies explicit control over the resolution of steep features, effectively avoiding UV runaways in ill-posed problems. On anti--diffusion and Burgers' shock formation, UVNG suppresses the high-frequency artifacts seen in the tested sequential baselines and approximately halves the relative error compared with the PINN baseline. It also produces stable evolutions of a nonlinear scalar model relevant to extensions of GR. The comparison with Israel--Stewart relaxation shows consistent later pulse propagation and nearly unchanged trajectories under the parameter variations examined, together with different splitting times during the early elliptic evolution. These properties position UVNG as a practical alternative for a broad variety of ill-posed PDEs and problems involving emerging discontinuities -- challenges that often arise in extensions to GR, where it offers a complementary way of regularizing the evolution. 

These results motivate several directions for future work. From a numerical perspective, systematic comparisons with PINN and other PDE solvers would help assess the computational efficiency and long-time accuracy of UVNG. Such assessments should account for its sequential-in-time nature: at fixed fitting budget per slice, the computational cost grows linearly with the number of time steps, while projection errors and ordinary time-discretization errors can accumulate across slices. A complementary direction is a more systematic assessment of regulator dependence, extending the parameter variations examined here. In our experiments, the penalty strength $\Upsilon$ is chosen separately for each problem, and its effect depends on both the fitting budget and the Fourier modes retained in the spectral correction. Studying these dependencies jointly would help guide the choice of regularization parameters. Finally, applications to most gravitational systems motivate extending the method beyond the one-dimensional problems with periodic boundary conditions presented here. This will require higher-dimensional implementations and adaptations of the spectral correction to accommodate outer boundaries or excision regions through additional constraints.


\begin{acknowledgments}
This work was supported in part by the Natural Sciences and Engineering Research Council
(NSERC) of Canada and the Simons Foundation through
Award SFI-MPS-BH-00012593-12 (LL). LL also acknowledges financial support through the Carlo Fidani Rainer Weiss Chair
at Perimeter Institute and CIFAR. HY is supported by the Natural Science Foundation of China (Grant 12573048). DL acknowledges support from the Beijing Major Science and Technology Project under Contract No. Z251100008125036. This research was supported in part by Perimeter Institute for
Theoretical Physics. Research at Perimeter Institute is supported in part by the Government of
Canada through the Department of Innovation, Science and Economic Development and by the
Province of Ontario through the Ministry of Colleges and Universities.
\end{acknowledgments}

\bibliography{references}

\newpage
\appendix
\begin{widetext}
\section{\label{appendix:proof} Cumulative deviation amplitude estimation for TENG}
\textbf{Euler step deviation estimation}\\
Let us denote the Euler step operator by $\mathcal{E}$:
\begin{equation*}
\mathcal{E} = \mathcal{I} + \Delta t \cdot\mathcal{L}
\end{equation*}
At each step $j$, the neural network approximation introduces an error $\epsilon^j(x)$, yielding the recurrence:
\begin{equation*}
u^j = \mathcal{E}\circ u^{j-1} + \epsilon^j
\end{equation*}
Unrolling this recurrence, the state at step $N$ can be expressed as:
\begin{equation*}
u^N = \mathcal{E}^N \circ u^0 + \sum_{j=1}^{N} \mathcal{E}^{N-j} \circ \epsilon^j
\end{equation*}
Here, $\mathcal{E}^N \circ u^0$ corresponds to the solution produced by the Euler scheme in the absence of approximation errors, while the second term represents the accumulated deviation. 

We decompose the errors $\{\epsilon^j\}_{j=1}^N$ into Fourier modes and isolate a single frequency $k$:
\begin{equation*}
\epsilon^j(x) = A_j e^{ikx}
\end{equation*}
Let $G(k) = 1+\Delta t \lambda k^{2n}$. 
With $\mathcal{W}=\sum_{m<2n} c_m\frac{\partial^m}{\partial x^m}$, we obtain: $\mathcal{W}\circ Ae^{ikx} = W(k)Ae^{ikx}$. Therefore:
\begin{equation*}
\mathcal{E}^{N-j}\circ\epsilon^j = \epsilon^j(G(k) + \Delta t W(k))^{N-j} = \epsilon^j G^{N-j}(k) + \mathcal{O}(A_j (N-j) G^{N-j-1}\Delta tW)
\end{equation*}
Therefore, for the $k$-th mode:
\begin{equation*}
\begin{split}
&u^N - \mathcal{E}^N \circ u^0 =  \sum_{j=1}^{N}\left[\epsilon^j G^{N-j}(k) + \mathcal{O}(A_j (N-j) G^{N-j-1}\Delta tW)\right] = [A_j=A\ \forall j] =\\&= Ae^{ikx} \sum_{m=0}^{N-1} G^{m}(k) + A\Delta tW \sum_{m=1}^{N-1} \mathcal{O}\left(m G^{m-1}(k)\right) = \\&= A\frac{G^{N}-1}{G-1}e^{ikx} + \mathcal{O}\left(A\Delta tW\frac{(N-1)G^N - NG^{N-1}+1}{(G-1)^2}\right) \approx \\ &\approx A\frac{G^{N}-1}{G-1}e^{ikx} + \mathcal{O}\left(A\Delta tW\frac{NG^{N-1}}{G-1}\right) =\\&= A\frac{(1+\Delta t \lambda k^{2n})^{N}-1}{\Delta t \lambda k^{2n}}e^{ikx} + \mathcal{O}\left(\frac{AN(1+\Delta t \lambda k^{2n})^{N-1}W(k)}{\lambda k^{2n}}\right)
\end{split}
\end{equation*}

\textbf{Heun step deviation estimation}\\
Let us denote the Heun step operator by $\mathcal{H}$:
\begin{equation*}
\mathcal{H} = \mathcal{I} + \frac{\Delta t}{2} \cdot\left(\mathcal{L} + \mathcal{L}\mathcal{E}\right)
\end{equation*}
At each step $j$, two approximation stages introduce two sources of deviation into the solution:
\begin{equation*}
\begin{split}
&u^j = \left[u^{j-1} + \frac{\Delta t}{2} \cdot\left(\mathcal{L}\circ u^{j-1} + \mathcal{L}\circ(\mathcal{E}\circ u^{j-1} + \epsilon_{\text{temp}}^j)\right)\right] + \epsilon_{\text{target}}^j =\\
&= \mathcal{H}\circ u^{j-1} + \frac{\Delta t}{2}\mathcal{L}\circ\epsilon_{\text{temp}}^j + \epsilon_{\text{target}}^j = \mathcal{H}\circ u^{j-1} + \epsilon_{\text{total}}^j
\end{split}
\end{equation*}
Here, $\epsilon_{\text{temp}}^j$ arises from the approximation of the temporary state:
\begin{equation*}
u_{\text{temp}}^j = \mathcal{E}\circ u^{j-1} + \epsilon_{\text{temp}}^j
\end{equation*}
and $\epsilon_{\text{target}}^j$ arises when fitting the final target, which includes the fitted predictor:
\begin{equation*}
u^j=u^{j-1}+\frac{\Delta t}{2}\left(\mathcal{L}\circ u^{j-1}+\mathcal{L}\circ u_{\text{temp}}^j\right)+\epsilon_{\text{target}}^j.
\end{equation*}
Given this recurrence, the state at step $N$ can be expressed as:
\begin{equation*}
u^N = \mathcal{H}^N \circ u^0 + \sum_{j=1}^{N} \mathcal{H}^{N-j} \circ \epsilon_{\text{total}}^j
\end{equation*}
We decompose both deviations $\{\epsilon_{\text{temp}}^j\}_{j=1}^N$ and $\{\epsilon_{\text{target}}^j\}_{j=1}^N$ into Fourier modes and isolate a single frequency $k$. The $k$-th mode of $\epsilon_{\text{total}}^j$ then takes the form:
\begin{equation*}
\epsilon_{\text{total}}^j = \left(\frac{\Delta t}{2}W(k)A_{\text{temp}}^j + \frac{\Delta t}{2} \lambda k^{2n}A_{\text{temp}}^j + A_{\text{target}}^j\right)e^{ikx} = A_{\text{total}}^j(k) e^{ikx}
\end{equation*}
Assuming that $\mathcal{W}$ consists of derivatives of order lower than $2n$, we obtain:
\begin{equation*}
\begin{split}
&\mathcal{H}\circ (A e^{ikx}) = A e^{ikx} \left(1 + \frac{\Delta t}{2}( W+\lambda k^{2n} + (W+\lambda k^{2n})(1+\Delta t \lambda k^{2n} + \Delta t W))\right) = \\
&=A e^{ikx} \left(\left[1 + \Delta t\lambda k^{2n} + \frac{\Delta t^2 \lambda^2 k^{4n}}{2}\right] + \mathcal{O}(\Delta t W + \Delta t^2 \lambda W k^{2n}) \right)
\end{split}
\end{equation*}
Let $\tilde G(k) = 1 + \Delta t\lambda k^{2n} + \frac{\Delta t^2 \lambda^2 k^{4n}}{2}$, then:
\begin{equation*}
\begin{split}
&\mathcal{H}^{N-j}\circ (A_{\text{total}}^j e^{ikx}) = \left(\tilde G(k) + \mathcal{O}(\Delta t W + \Delta t^2 \lambda W k^{2n})\right)^{N-j} A_{\text{total}}^j e^{ikx} =\\ &= A_{\text{total}}^j e^{ikx} \cdot \tilde G^{N-j}(k) + \mathcal{O}(A_{\text{total}}^j (N-j) \tilde G^{N-j-1} (\Delta t W + \Delta t^2 \lambda W k^{2n}))
\end{split}
\end{equation*}
Therefore, the deviation of the $k$-th mode from the reference solution is given by:
\begin{equation*}
\begin{split}
&u^N - \mathcal{H}^N \circ u^0 = \sum_{j=1}^{N} \mathcal{H}^{N-j} \circ \epsilon_{\text{total}}^j = 
\sum_{j=1}^{N}\left[\epsilon_{\text{total}}^j \tilde G^{N-j}(k) + \mathcal{O}(A_{\text{total}}^j (N-j) \tilde G^{N-j-1} (\Delta t W + \Delta t^2 \lambda W k^{2n}))\right] = \\ &= \left[A_{\text{temp}}^j=A_{\text{target}}^j=A\ \forall j\implies A_{\text{total}}^j = A(1 + \frac{\Delta t}{2}W(k) + \frac{\Delta t}{2} \lambda k^{2n})\equiv A_{\text{total}}(k)\ \forall j\right] =\\ &= A_{\text{total}}\cdot e^{ikx}\sum_{j=1}^{N} \tilde G^{N-j}(k) + A_{\text{total}}(\Delta t W + \Delta t^2 \lambda W k^{2n})\sum_{j=1}^{N} \mathcal{O}\left((N-j)\tilde G^{N-j-1}(k)\right) = \\ &= A_{\text{total}}\cdot e^{ikx} \sum_{m=0}^{N-1} \tilde G^{m}(k) + A_{\text{total}}(\Delta t W + \Delta t^2 \lambda W k^{2n})\sum_{m=1}^{N-1} \mathcal{O}\left(m\tilde G^{m-1}(k)\right) = \\ &= A_{\text{total}} \frac{\tilde G^N - 1}{\tilde G - 1} e^{ikx} + \mathcal{O}\left(A_{\text{total}}(\Delta t W + \Delta t^2 \lambda W k^{2n})\frac{(N-1)\tilde G^N - N\tilde G^{N-1}+1}{(\tilde G-1)^2}\right) = \\ & = A_{\text{total}} \frac{\tilde G^N - 1}{\tilde G - 1} e^{ikx} + \mathcal{O}\left(A_{\text{total}}(\Delta t W + \Delta t^2 \lambda W k^{2n})\frac{N\tilde G^{N-1}}{\tilde G-1}\right) = \\ &= A_{\text{total}}\frac{\left(1 + \Delta t\lambda k^{2n} + \frac{\Delta t^2 \lambda^2 k^{4n}}{2}\right)^N - 1}{\Delta t\lambda k^{2n} + \frac{\Delta t^2 \lambda^2 k^{4n}}{2}}e^{ikx} + \mathcal{O}\left(\frac{A_{\text{total}} N W(k)}{\lambda k^{2n}}\left(1 + \Delta t\lambda k^{2n} + \frac{\Delta t^2 \lambda^2 k^{4n}}{2}\right)^{N-1}\right), \\ & \text{where}\ A_{\text{total}} = A\left(1 + \frac{\Delta t}{2} (W(k) + \lambda k^{2n})\right).
\end{split}
\end{equation*}

\section{\label{appendix:2D_runaway} UV runaway in higher dimensions}
Substituting the leading amplification factors~\eqref{eq:Euler_new_G} and \eqref{eq:Heun_new_G} into the proof framework of Appendix \ref{appendix:proof}, we obtain deviation estimates for the multidimensional case. Specifically, for any wave vector $\boldsymbol{k}$ satisfying $\sum_{|p|=2n} \lambda_{p}(k_1^{p_1}\dots k_d^{p_d})>0$, the cumulative error amplitude at step $N$ is asymptotically given by:
\begin{equation}
\begin{aligned}
\mathcal{A}_{\text{Euler}}(N,\boldsymbol{k}) &=
A\frac{\left(1+\Delta t \sum_{|p|=2n} \lambda_{p}(k_1^{p_1}\dots k_d^{p_d})\right)^{N}-1}{\Delta t \sum_{|p|=2n} \lambda_{p}(k_1^{p_1}\dots k_d^{p_d})} +\\
&+\mathcal{O}\left(\frac{AN\left(1+\Delta t \sum_{|p|=2n} \lambda_{p}(k_1^{p_1}\dots k_d^{p_d})\right)^{N-1}W(\boldsymbol{k})}{\sum_{|p|=2n} \lambda_{p}(k_1^{p_1}\dots k_d^{p_d})}\right)
\end{aligned}
\end{equation}
\begin{equation}
\begin{aligned}
\mathcal{A}_{\text{Heun}}&(N,\boldsymbol{k}) = A_{\text{total}}\frac{\left(1 + \Delta t\sum_{|p|=2n} \lambda_{p}(k_1^{p_1}\dots k_d^{p_d}) + \frac{\Delta t^2 }{2}\left(\sum_{|p|=2n} \lambda_{p}(k_1^{p_1}\dots k_d^{p_d})\right)^2\right)^N - 1}{\Delta t\sum_{|p|=2n} \lambda_{p}(k_1^{p_1}\dots k_d^{p_d}) + \frac{\Delta t^2 }{2}\left(\sum_{|p|=2n} \lambda_{p}(k_1^{p_1}\dots k_d^{p_d})\right)^2} + \\
&+ \mathcal{O}\left(\frac{A_{\text{total}} N W(\boldsymbol{k})}{\sum_{|p|=2n} \lambda_{p}(k_1^{p_1}\dots k_d^{p_d})}\left(1 + \Delta t\sum_{|p|=2n} \lambda_{p}(k_1^{p_1}\dots k_d^{p_d}) + \frac{\Delta t^2 }{2}\left(\sum_{|p|=2n} \lambda_{p}(k_1^{p_1}\dots k_d^{p_d})\right)^2\right)^{N-1}\right),\\
&\text{where}\ A_{\text{total}} = A\left(1 + \frac{\Delta t}{2} \left(W(\boldsymbol{k}) + \sum_{|p|=2n} \lambda_{p}(k_1^{p_1}\dots k_d^{p_d})\right)\right).
\end{aligned}
\end{equation}

It is noteworthy that, unlike the 1D setting where higher values of $k$ monotonically increase amplitude growth, in multidimensional case the growth is governed by the polynomial form $\sum_{|p|=2n} \lambda_{p}k_1^{p_1}\dots k_d^{p_d}$ which is parametrized by the set of coefficients $\{\lambda_p\}_{|p|=2n}$. Consequently, the orientation of the wave vector $\boldsymbol{k}$ plays a decisive role. For example, if the value of $\lambda_p$ corresponding to the multi-index $p=(2n,\boldsymbol{0})$ is zero, the highest-order contribution will vanish along $\boldsymbol{k}=(\tilde{k},\boldsymbol{0})$.

\section{\label{appendix:interpretation} Regularization interpretation}
Let $\mathcal{X}=[x_0,x_1]$ be a one-dimensional periodic domain equipped with the mean-normalized inner product:
\begin{equation*}
\langle f,g\rangle = \frac{1}{|\mathcal{X}|}\int_{\mathcal{X}}f(x)g(x)dx,\quad \langle g,g\rangle \equiv \left\|g\right\|^2
\end{equation*}
Consider the regularization functional:
\begin{equation*}
\Omega(u) = \left\|\frac{\partial^n u}{\partial x^n}\right\|^2
\end{equation*}
To determine its functional derivative, let us consider a perturbation $u\rightarrow u+\varepsilon h$, where $h$ is an arbitrary smooth periodic function. The first variation of $\Omega$ is
\begin{equation*}
\delta\Omega(u,h) =  \left.\frac{d}{d\varepsilon}\Omega(u+\varepsilon h)\right|_{\varepsilon=0} = \frac{2}{|\mathcal{X}|} \int_{\mathcal{X}} \frac{\partial^n u}{\partial x^n}\frac{\partial^n h}{\partial x^n} dx
= \frac{2(-1)^n}{|\mathcal{X}|} \int_{\mathcal{X}} \frac{\partial^{2n} u}{\partial x^{2n}} h dx,
\end{equation*}
where the last equality follows from repeated integration by parts, with all boundary contributions vanishing due to periodicity $\left.\frac{\partial^k u}{\partial x^k}\right|_{x=x_0}=\left.\frac{\partial^k u}{\partial x^k}\right|_{x=x_1}\ \forall k\leq2n$.
Consequently, the functional derivative is:
\begin{equation*}
\frac{\delta \Omega}{\delta u} = 2 (-1)^n \frac{\partial^{2n} u}{\partial x^{2n}}
\end{equation*}
Let us consider the update $\Delta u = \Delta t \mathcal{L}\circ u$. Without parameter-increment regularization ($\gamma=0$), the linear system is
\begin{equation*}
\begin{pmatrix} J \\ \Upsilon \nabla_\theta \Omega(u_\theta)^T \end{pmatrix} \Delta \theta = \begin{pmatrix} \Delta u \\ 0 \end{pmatrix}
\end{equation*}
In the collocation-to-continuum approximation, applying the chain rule gives:
\begin{equation*}
\nabla_\theta \Omega(u_\theta)^T\cdot\Delta\theta = \langle\frac{\delta\Omega}{\delta u}, J\cdot\Delta\theta\rangle
\end{equation*}
And the optimization problem takes the form:
\begin{equation*}
\Delta \theta = \arg \min_{\Delta \theta} \left(M\right),\quad\text{where}\ M=\left\| J\cdot\Delta\theta - \Delta u \right\|^2 + \frac{\Upsilon^2}{s}\langle\frac{\delta\Omega}{\delta u}, J\cdot\Delta\theta\rangle^2.
\end{equation*}
Let us denote $w = J\cdot\Delta\theta$, $\Upsilon_{\text{eff}}=\frac{\Upsilon}{\sqrt{s}}$ and find an analytical solution of the optimization problem (we consider the idealized limit in which the neural-network tangent space spans the function space, allowing to minimize $M$ with respect to $w$).
\begin{equation*}
\frac{\delta M}{\delta w} = 2(w - \Delta u) + 2\Upsilon_{\text{eff}}^2 \langle\frac{\delta\Omega}{\delta u}, w\rangle \frac{\delta\Omega}{\delta u}
\end{equation*}
Notice that:
\begin{equation*}
w = \Delta u - \frac{\Upsilon_{\text{eff}}^2 \langle\frac{\delta\Omega}{\delta u},\Delta u\rangle}{1 + \Upsilon_{\text{eff}}^2 \left\|\frac{\delta\Omega}{\delta u}\right\|^2} \frac{\delta\Omega}{\delta u}
\end{equation*}
is a solution to the problem $\frac{\delta M}{\delta w} = 0$. Therefore (assuming only one update step is performed):
\begin{equation*}
u_{\theta_{t+\Delta t}} - u_{\theta_{t}} = w + \mathcal{O}(\|\Delta\theta\|^2),\qquad w = \Delta t \left(\mathcal{L}\circ u - \frac{\Upsilon_{\text{eff}}^2 \langle\frac{\delta\Omega}{\delta u},\mathcal{L}\circ u\rangle}{1 + \Upsilon_{\text{eff}}^2 \left\|\frac{\delta\Omega}{\delta u}\right\|^2} \frac{\delta\Omega}{\delta u}\right)
\end{equation*}
Assuming a regular parameter update with $\Delta\theta=\mathcal{O}(\Delta t)$, the nonlinear remainder vanishes after division by $\Delta t$. Formally taking $\Delta t\to0$ and substituting $\delta\Omega/\delta u$ gives:
\begin{equation*}
\frac{\partial u}{\partial t} = \mathcal{L}\circ u - \nu(u) \frac{\partial^{2n} u}{\partial x^{2n}},\quad\text{where}\ \nu(u)=\frac{4\Upsilon_{\text{eff}}^2 \langle\frac{\partial^{2n}u}{\partial x^{2n}},\mathcal{L}\circ u\rangle}{1 + 4\Upsilon_{\text{eff}}^2 \left\|\frac{\partial^{2n}u}{\partial x^{2n}}\right\|^2}.
\end{equation*}
\end{widetext}

\section{\label{appendix:ablation} Ablation study}
\subsection{\label{appendix:step_ablation} Correction steps}
We compare the correction steps on fixed target functions. All models share the same architecture, training parameters, and initial states; the only variation is the update applied every tenth iteration: a standard UVNG projection, a linear correction, or a spectral correction. For target functions, we consider:
\begin{itemize}
    \item Continuous function with a finite Fourier spectrum (multi-frequency signal):
    \begin{equation*}
    u_{\text{target}}(x) = \sum_i A_i \sin(k_i x+\phi_i),
    \end{equation*}
    where $\forall i\ A_i\in[-1;1],\ k_i\in[-5;5],\ \phi_i \in [-\pi;\pi]$ are randomly chosen coefficients.
    
    \item Continuous function with an infinite Fourier spectrum (Gaussian function):
    \begin{equation*}
    u_{\text{target}}(x) = e^{-2x^2}
    \end{equation*}
    \item Discontinuous function (Heaviside step function):
    \begin{equation*}
    u_{\text{target}} = \theta(x) = \left\{
    \begin{aligned}
        &0,\quad \text{if}\ x<0 \\
        &\frac{1}{2},\quad \text{if}\ x = 0 \\
        &1,\quad \text{if}\ x > 0
    \end{aligned}\right.
    \end{equation*}
\end{itemize}

\begin{table*}
  \caption{Approximation accuracy $\mathrm{E}_{\text{rel}}$ for fixed target functions, using an MLP with architecture [1, 20, 20, 20, 1] after 5000 iterations (about 3 min of training in each case).}
  \label{tab:ablation_corrstep_1}
  \centering
  \begin{ruledtabular}
  \begin{tabular}{lccccc}
     & \textbf{no corr} & \multicolumn{3}{c}{\textbf{spectral corr}} & \textbf{linear corr}\\
    \textbf{Function} \textbackslash\ \textbf{Method} & & \textbf{$d=3$} & \textbf{$d=5$} & \textbf{$d=7$} & \\ \hline
    Multi-frequency signal & $6.9\times10^{-3}$ & $1.9\times10^{-3}$ & $3.4\times10^{-4}$ & $1.2\times10^{-4}$ & $2.0\times10^{-6}$\\
    Gauss function & $6.4\times10^{-3}$ & $6.7\times10^{-4}$ & $2.6\times10^{-4}$ & $1.4\times10^{-4}$ & $1.2\times10^{-7}$\\
    Heaviside function & $8.0\times10^{-2}$ & $8.0\times10^{-2}$ & $8.0\times10^{-2}$ & $1.2\times10^{-1}$ & $2.0\times10^{-1}$\\
  \end{tabular}
  \end{ruledtabular}
\end{table*}
For the smooth targets in Table~\ref{tab:ablation_corrstep_1}, spectral correction improves approximation accuracy as more modes are retained, and unrestricted linear correction gives still smaller errors. Neither correction improves the Heaviside fit at the budget tested. Improved fitting can also recover unwanted high-frequency content that we seek to avoid. To study this trade-off, we fit signals supplemented with artificial noise:\\
$u_{\mathrm{truth}}(x)$ is the original multi-frequency, Gaussian, or Heaviside signal;\\
$\delta(x)=\sum_i A_i\sin(k_i x+\phi_i)$ is the noise, with randomly chosen coefficients $A_i\in[10^{-5},10^{-3}]$, $k_i\in[9,16]$, and $\phi_i\in[-\pi,\pi]$;\\
$u_{\mathrm{target}}(x)=u_{\mathrm{truth}}(x)+\delta(x)$ is the function supplied to the approximator.

To measure the network's ability to approximate the true signal while actively resisting the noise, we introduce the noise approximation relation $\eta$:
\begin{equation*}
\eta = \frac{\mathrm{E}_{\text{noise}}}{\mathrm{E}_{\text{signal}}}
\end{equation*}
where the respective $L^2$ errors are defined as:
\begin{equation*}
\begin{split}
&\mathrm{E}_{\text{signal}} = \frac{\|u_{\text{pred}}-u_{\text{truth}}\|_{L^2}}{\|u_{\text{truth}}\|_{L^2}},\\
&\mathrm{E}_{\text{noise}} = \frac{\|(u_{\text{pred}}-u_{\text{truth}})-\delta\|_{L^2}}{\|\delta\|_{L^2}}
\end{split}
\end{equation*}
The ratio $\eta$ compares the relative error in fitting the noise with the relative error in fitting the underlying signal. A lower value of $\eta$ indicates that the neural network is fitting the noisy high-frequency features rather than filtering them out. Table~\ref{tab:ablation_corrstep_2} illustrates that, for smooth targets, retaining sufficiently many modes in spectral correction or applying linear correction reduces resistance to fitting high-frequency noise.

\begin{table*}
  \caption{Relation $\eta$ in different settings.}
  \label{tab:ablation_corrstep_2}
  \centering
  \begin{ruledtabular}
  \begin{tabular}{lccccc}
     & \textbf{no corr} & \multicolumn{3}{c}{\textbf{spectral corr}} & \textbf{linear corr}\\
    \textbf{Function} \textbackslash\ \textbf{Method} & & \textbf{$d=3$} & \textbf{$d=5$} & \textbf{$d=7$} & \\ \hline
    Multi-frequency signal & 882 & 1015 & 1030 & 861 & 694\\
    Gauss function & 528 & 1130 & 1050 & 589 & 513\\
    Heaviside function & 769 & 769 & 769 & 770 & 770\\
  \end{tabular}
  \end{ruledtabular}
\end{table*}

\subsection{\label{appendix:AD_ablation} UV runaway}
\begin{figure*}
  \centering
  \includegraphics[width=0.8\linewidth]{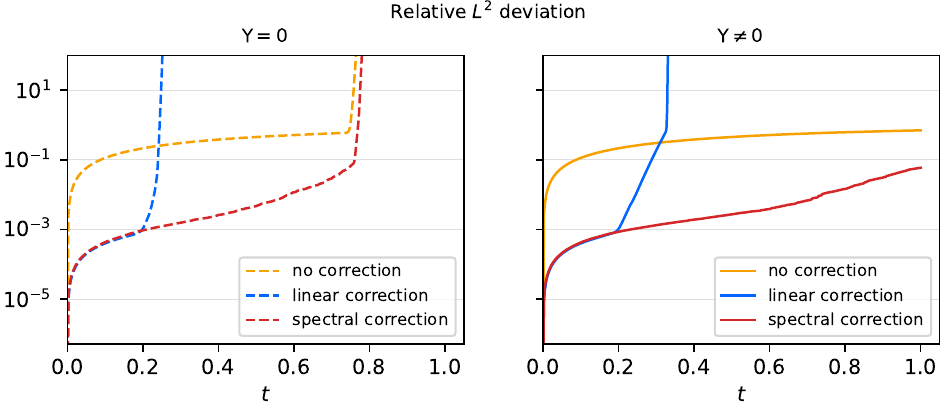}
  \caption{Temporal evolution of the relative $L^2$ error for different UVNG settings for problem \eqref{eq:AD}. \textbf{(Left)} Without derivative regularization. \textbf{(Right)} With derivative regularization.}
  \label{fig:AD_ablation}
\end{figure*}

To assess the individual contribution of each component in the UVNG framework -- specifically, derivative regularization and correction steps, we evaluate different configuration settings on the anti--diffusion equation \eqref{eq:AD}. The temporal evolution of the relative $L^2$ error across these configurations is presented in Fig.~\ref{fig:AD_ablation}.\\
The empirical conclusions drawn from Fig.~\ref{fig:AD_ablation} align with the established mathematical intuition:
\begin{itemize}
\item Without regularization of the mean squared derivative, all shown solutions develop UV runaway.
\item Linear correction improves accuracy but inherently triggers UV runaway, regardless of whether regularization is applied.
\item Combining regularization with spectral correction gives the smallest final error and maintains stability over the considered interval.
\end{itemize}
These results support using the regularization of mean squared derivative together with spectral correction.

\section{\label{appendix:reference_solutions} Reference solutions}
\textbf{Anti--diffusion equation}\\
An analytical solution to problem \eqref{eq:AD} is given by:

\begin{equation*}
u(x,t) = A\sin(kx)e^{k^2\lambda t}
\end{equation*}

\textbf{Burgers equation (inviscid limit)}\\
The reference solution to problem \eqref{eq:Burgers} prior to complete shock formation at $t_c=\frac{\sqrt{e}}{2}$ is obtained by the method of characteristics, while for $t\ge t_c$ we utilize a high-resolution finite-difference solution of the Burgers’ equation with small viscosity $\nu=5\cdot10^{-4}$ (spatial discretization utilizes central differences on a fine grid ($N=5000$), evolved via RK4 integration).

\section{\label{appendix:exp_setup} Experimental setup}
\subsection{UVNG}
In all experiments, we use Euler stepping and a separate MLP for each evolved variable, with architecture [1, 20, 20, 20, 1] and activation function $\sin(x)$. Boundary conditions are imposed through additional constraints in the linear system~\eqref{formula:UVNG_system}. Each fit uses $s=1000$ collocation points, parameter subsamples of size $p=30$, and $\gamma=0.01$. The remaining hyperparameters are problem-specific:
\begin{itemize}
\item For the problem \eqref{eq:AD}, we use $\Delta t = 0.002$, $\Upsilon=0.08$. Training settings: $N_g=50$, $N_\ell=9$ (500 iterations in total, with spectral correction applied every 10th iteration for $d=5$ modes).
\item For the problem \eqref{eq:Burgers}, we use $\Delta t = 0.005$, $\Upsilon=0.01$. Training settings: $N_g=100$, $N_\ell=49$ (5000 iterations in total, with spectral correction applied every 50th iteration for $d=5$ modes).
\item For the problem \eqref{eq:EFT}, we use $\Delta t = 0.02$, $\Upsilon=1.00$. Training settings: 3000 iterations in total. These settings are identical for all variables.
\end{itemize}

We choose $\Upsilon$ empirically to suppress visible UV growth over the interval of interest without an unnecessarily large fitting bias. For a fixed fitting budget and mode selection, one practical starting point is the smallest penalty that meets this criterion. When the number of collocation points is changed, the normalization $\Upsilon_{\mathrm{eff}}=\Upsilon/\sqrt{s}$ must also be taken into account.\\
Linear systems are solved using \texttt{torch.linalg.lstsq} on GPU. Depending on the system dimensions, the underlying driver utilizes QR factorization for overdetermined cases and LQ factorization for underdetermined cases.

\subsection{PINN}
The PINN baseline uses causal loss weighting~\cite{Perdikaris2024} and residual-based adaptive sampling. For the causal loss, the temporal domain is divided into half-open intervals $[t_k,t_{k+1})$, and the total residual loss is computed as a dynamically weighted sum:
\begin{equation*}
\mathcal{L}_{\text{res}} = \frac{1}{\sum_{k=1}^{N_t} w_k} \sum_{k=1}^{N_t} w_k \mathcal{L}_{\text{res}}([t_k,t_{k+1})),
\end{equation*}
where the residual on a specific interval $[t_k,t_{k+1})$ is evaluated over $N_k$ collocation points:
\begin{equation*}
\mathcal{L}_{\text{res}}([t_k,t_{k+1})) = \frac{1}{N_k} \sum_{j=1}^{N_k}  \textit{Res}^2(u_{\theta}(t_j, x_j)).
\end{equation*}
The causal weights $w_k$ are defined sequentially as:
\begin{equation*}
w_k = \exp\left(-\varepsilon \sum_{i=1}^{k-1} \mathcal{L}_{\text{res}}([t_i,t_{i+1}))\right),
\end{equation*}
where $\varepsilon$ is a strictly positive hyperparameter. These weights downweight later intervals while the residual on preceding intervals remains large; they encourage temporal ordering rather than imposing a hard causal constraint.

To focus representational capacity on high-error regions, collocation points are resampled based on the normalized point-wise residual $\tilde{r}(x)$:
\begin{equation*}
\begin{split}
&p(t,x) \propto \tilde{r}(t, x) + \delta,\\
\text{where} &\ \tilde{r}(t, x) = \frac{|\textit{Res}(u_{\theta}(t, x))| - \min |\textit{Res}(u_{\theta}(t, x))|}{\max |\textit{Res}(u_{\theta}(t, x))| - \min |\textit{Res}(u_{\theta}(t, x))|}
\end{split}
\end{equation*}
Here, $\delta$ ensures a non-zero probability of sampling anywhere in the domain.

In our experiments, we use $\varepsilon=0.1$, $N_t=30$ for causal training and $\delta = 0.005$ for residual-based generation of $N=10^4$ points. The network architecture is [2, 50, 50, 50, 1] with $\sin(x)$ activation. Optimization is performed using the ADAM optimizer with an exponentially decaying learning rate for $10^5$ iterations.

\subsection{Other baseline methods}
Across the benchmark problems we also consider other time-sequential approaches in order to investigate how UV runaway occurs in different scenarios:
\begin{itemize}
\item \textbf{FD scheme for problem \eqref{eq:AD}:} For the anti--diffusion equation, we implement Crank-Nicolson scheme using a central difference discretization for the spatial operator $\mathcal{L}$ under periodic boundary conditions. The discrete update is formulated as:
\begin{equation}
(I + \alpha D_2) \mathbf{u}^{n+1} = (I - \alpha D_2) \mathbf{u}^n
\end{equation}
where $\mathbf{u}^n$ is the solution vector at time step $n$, $I$ is the identity matrix, and $D_2$ is the discrete second-order derivative operator with periodic wrap-around. The parameter $\alpha = \frac{\lambda \Delta t}{2 \Delta x^2}$ represents the mesh Fourier number. While the scheme is stable for well-posed diffusive problems ($\lambda < 0$), it provides no damping for high-frequency modes in the ill-posed anti-diffusive regime.

\item \textbf{FD scheme for problem \eqref{eq:Burgers}:} For Burgers' equation, we use RK4 time integration and fourth-order centered differences for the flux derivative, with $f(u)=u^2/2$ and periodic boundary conditions. This baseline includes neither viscosity nor a shock-capturing flux.

\item \textbf{FD scheme for problem \eqref{eq:EFT_fixed}:} For the model problem stabilized by the Israel--Stewart approach, we implement the same scheme as suggested in \cite{Allwright2019}: a third-order Runge-Kutta (SSP-RK3) method for time integration, combined with a fourth-order centered finite-difference discretization for the spatial derivatives under periodic boundary conditions. The time step is chosen as $\Delta t=\Delta x/4$ in accordance with the CFL condition.

\item \textbf{FD scheme for problem \eqref{eq:EFT} with higher-order correction:} We consider the problem \eqref{eq:EFT} and augment the evolution equation for $g$ with the fourth-order correction $-\nu(g)\frac{\partial^4g}{\partial x^4}$, where the solution-dependent coefficient $\nu(g)$ is computed at each stage according to Proposition~\ref{proposition-2}. We set $\Upsilon_{\text{eff}}=\frac{\Upsilon}{\sqrt{s}}$, where $\Upsilon=1$ and $s=1000$ are, respectively, the regularization parameter and the number of collocation points used in the corresponding UVNG experiment. For the numerical implementation, we consider the same numerical scheme as for the Israel--Stewart stabilized system. The fourth-order derivative is approximated using a fourth-order accurate centered finite-difference stencil. Unlike the previous scheme, the time step is chosen adaptively as $\Delta t=\min(\frac{\Delta x}{4},\ \frac{\Delta x^4}{10|\nu(g)|})$, with the second restriction omitted when $\nu(g)=0$.

\item \textbf{TENG:} Our TENG baseline includes parameter-increment regularization ($\gamma=0.01$) and a linear correction every 50 fitting iterations.  While these additions improve stability and accuracy in well-posed scenarios, they do not fundamentally address the UV runaway. All TENG experiments in Sec.~\ref{experiments} use Euler time integration. The Heun run in Fig.~\ref{fig:deviation_bounds} separately tests the accumulation mechanism in Proposition~\ref{proposition}.

\item \textbf{OBTI:} As a baseline sequential-in-time method, we also evaluate the OBTI framework. The key distinction from TENG lies in the chosen optimization landscape. While TENG exploits the convex squared-error objective in function space ($u$-space) before projecting the update, OBTI attempts to find the parameters for the subsequent slice, $\theta_{t+\Delta t}$, by minimizing a loss function directly in the highly non-convex parameter space ($\theta$-space) using established optimizers (ADAM, L-BFGS). Formally, given the same explicit target $u_{\text{target}}(x)$ (obtained with Euler scheme in our experiments), the next step is defined entirely by this direct minimization:

\begin{equation}
\theta_{t+\Delta t} = \arg \min_{\theta } \textit{L}(u_{\theta}, u_{\text{target}})
\end{equation}

While this approach avoids a separate projection step, it faces significant practical challenges. The non-convexity of the $\theta$-space landscape requires many solver iterations, leading to high computational costs and generally lower optimization accuracy. Consequently, OBTI is often outperformed by TENG in well-posed scenarios \citep{Luo2024}. However, when dealing with ill-posed problems, there was an initial aspiration that OBTI's direct parameter optimization might avoid the amplification of high-frequency errors associated with the spectral bias phenomenon.

To further stabilize the method against such amplification, we also explored augmenting the OBTI objective function with regularization terms analogous to those utilized in UVNG:
\begin{equation*}
    \textit{L}_{\text{reg}} = \textit{L}(u_{\theta}, u_{\text{target}}) + \gamma \|\theta\|_2^2 + \Upsilon \left\| \frac{\partial^n u_{\theta}}{\partial x^n}(x)\right\|_{L^2(\mathcal{X})}^2
\end{equation*}
where $\gamma$ scaled the weight decay and $\Upsilon$ penalized the $n$-th order spatial derivative. Interestingly, this formulation does not yield stable behavior. Because the penalty minimizes the absolute magnitude of the mean squared derivative itself -- rather than the incremental change in the derivative, as is done in UVNG, all features of the solution are rapidly erased over successive time steps. For this reason, only the original OBTI method is considered.

\end{itemize}
For all time-sequential methods in the problem \eqref{eq:AD}, we use a time step of $\Delta t = 0.002$, while for the problem \eqref{eq:Burgers} $\Delta t = 0.005$.

\section{\label{appendix:Burgers} Burgers' equation further study}
\subsection{\label{appendix:Burgers-shock} Shock propagation}
\begin{figure}[tbp]
  \centering
  \includegraphics[width=0.8\linewidth]{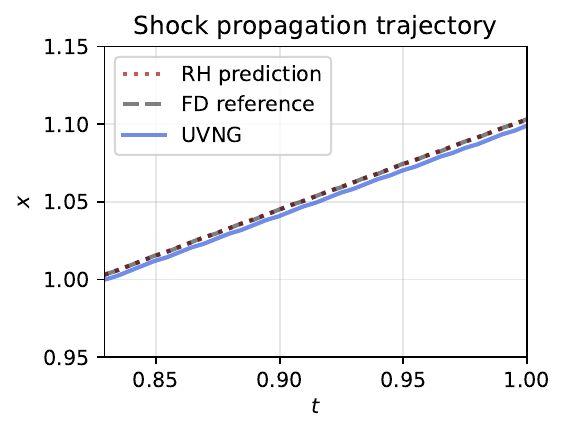}
  \caption{Shock propagation trajectory for the Burgers' equation \eqref{eq:Burgers} beyond the critical time $t_c$. The shock positions extracted from UVNG and the viscous finite-difference reference solution are compared with the Rankine-Hugoniot (RH) prediction.}
  \label{fig:burgers_st}
\end{figure}

\begin{figure}[tbp]
  \centering
  \includegraphics[width=0.8\linewidth]{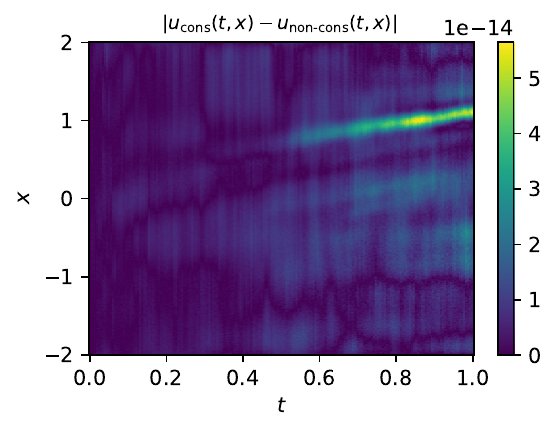}
  \caption{Absolute difference between UVNG solutions obtained using the conservative form $\frac{1}{2}(u^2)_x$ and the non-conservative form $u u_x$ for the inviscid Burgers' equation.}
  \label{fig:burgers_adv}
\end{figure}

To assess the front motion beyond shock formation, we compare the shock trajectories obtained with UVNG and the reference solution against the Rankine-Hugoniot condition. For the Burgers equation, the shock propagation speed is given by:
\begin{equation*}
s_{\mathrm{RH}}(t) = \frac{u_L(t)+u_R(t)}{2},
\end{equation*}
where \(u_L\) and \(u_R\) denote the solution values immediately to the left and right of the shock, respectively. We estimate these states from the viscous reference solution by extrapolating the solution profiles from outside the shock transition layer. The corresponding Rankine-Hugoniot trajectory is obtained by integrating this speed, initialized at the analytically determined shock formation point \((t_c,x_c)=(\frac{\sqrt{e}}{2},1)\). Fig.~\ref{fig:burgers_st} compares the resulting trajectory with the shock positions extracted from the UVNG and reference solutions. The close agreement confirms that UVNG reproduces the expected macroscopic shock propagation despite the finite width of its resolved shock.\\

\subsection{\label{appendix:Burgers-advection} Advection term formulation}
In the context of classical grid-based numerical methods, simulating the inviscid Burgers' equation is highly sensitive to the algebraic formulation of the advection term. Specifically, solving the non-conservative form $u u_x$ rather than the conservative form $\frac{1}{2}(u^2)_x$ typically leads to incorrect shock propagation speeds \citep{toro2009riemann}.
However, experiments with UVNG demonstrate no meaningful discrepancy in performance between these two formulations (see Fig.~\ref{fig:burgers_adv}). This robustness is likely due to the use of automatic differentiation. Unlike finite difference stencils, which suffer from distinct truncation errors depending on the algebraic arrangement, AD computes exact point-wise derivatives of the computational graph. By the exactness of the analytical chain rule, the operations $u_{\theta} \partial_x u_{\theta}$ and $\frac{1}{2}\partial_x(u_{\theta}^2)$ are identical down to floating-point precision for any differentiable function $u_{\theta}$. Consequently, the resulting values remain the same, thus making UVNG unreceptive to the conservative versus non-conservative formulation.

\end{document}